\documentclass[aps,prd,superscriptaddress,nofootinbib,secnumarabic,preprintnumbers]{revtex4-2}[10pt]
\pdfoutput=1
\usepackage{latexsym}
\usepackage{ucs}
\usepackage{mathrsfs, amsmath, amssymb}
\usepackage{diagbox, slashed, makecell}
\usepackage{hyperref, url}
\usepackage{color}
\usepackage{graphicx}
\usepackage{appendix}
\usepackage{multirow}
\usepackage{enumitem}
\usepackage{indentfirst, layout, geometry}

\newcommand{\mr}{\mathrm}
\newcommand{\mc}{\mathcal}

\graphicspath{{figures/}}

\begin{document}

\begin{sloppypar}

\title{A leptoquark and vector-like quark extended model for the simultaneous explanation of the $W$ boson mass and muon $g-2$ anomalies}


\author{Shi-Ping He}
\email{shiping.he@apctp.org}
\affiliation{Asia Pacific Center for Theoretical Physics, Pohang 37673, Korea}

\date{\today}

\begin{abstract}
The CDF collaboration recently announced a new measurement result of the $W$ boson mass, and it is in tension with the standard model (SM) prediction. In this paper, we explain this anomaly in the vector-like quark (VLQ) $(X,T,B)_{L,R}$ and leptoquark (LQ) $S_3$ extended model. In this model, both the VLQ and LQ have positive corrections to the $W$ boson mass. Moreover, it can also be a solution to the $(g-2)_{\mu}$ anomaly because of the chiral enhancements from top, $T$, and $B$ quarks.
\end{abstract}

\maketitle

\clearpage

\section{Introduction}
The SM has provided powerful description of elementary particle physics, and it can explain most of the experiments with high precision \cite{ParticleDataGroup:2020ssz}. However, it is challenged by some experiments these years, for example $(g-2)_{\mu}$ \cite{Muong-2:2006rrc, Muong-2:2021ojo} and $B$ decay anomalies \cite{London:2021lfn}. Recently, the CDF collaboration announced a measurement of $W$ boson mass \cite{CDF:2022hxs}, which shows the seven standard deviations heavier than the SM prediction. On the one hand, we still need to scrutinize the theoretical and experimental uncertainties \cite{Bacchetta:2018lna}. On the other hand, this may be a signature of new physics \cite{Campagnari:2022}. Currently, there are many studies dedicated to explaining this CDF anomaly \cite{Fan:2022dck, *Lu:2022bgw, *Du:2022pbp, *Tang:2022pxh, *Yang:2022gvz, *Strumia:2022qkt, *Blennow:2022yfm, *Cacciapaglia:2022xih, *Liu:2022jdq, *Arias-Aragon:2022ats, *Fan:2022yly, *Sakurai:2022hwh, *Athron:2022isz, *Asadi:2022xiy, *Song:2022xts, *Bahl:2022xzi, *Cheng:2022jyi, *Lee:2022nqz, *Heckman:2022the, *Bagnaschi:2022whn, *DiLuzio:2022xns, *Babu:2022pdn, *Paul:2022dds, *Biekotter:2022abc, *Cheung:2022zsb, *Du:2022brr, *Endo:2022kiw, *Crivellin:2022fdf, *Heo:2022dey, *Ahn:2022xax, *Zheng:2022irz, *FileviezPerez:2022lxp, *Ghoshal:2022vzo, *Peli:2022ybi, *Kawamura:2022uft, *Kanemura:2022ahw, *Nagao:2022oin, *Mondal:2022xdy, *Zhang:2022nnh, *Carpenter:2022oyg, *Popov:2022ldh, *Arcadi:2022dmt, *Chowdhury:2022moc, *Borah:2022obi, *Cirigliano:2022qdm, *Zeng:2022lkk, *Du:2022fqv, *Ghorbani:2022vtv, *Bhaskar:2022vgk, *Baek:2022agi, *Borah:2022zim, *Lee:2022gyf, *Almeida:2022lcs, *Cheng:2022aau, *Addazi:2022fbj, *Heeck:2022fvl, *Abouabid:2022lpg, *TranTan:2022kpq, *Batra:2022pej, *Benbrik:2022dja, *Cai:2022cti, *Zhou:2022cql, *Chen:2022ocr, *Gupta:2022lrt, *Botella:2022rte, *Wang:2022dte, *Basiouris:2022wei}.

To explain the $W$ mass anomaly, we can introduce new fermions and scalars. The VLQs are well motivated in many new physics models, for example, composite Higgs models \cite{Agashe:2004rs, Panico:2015jxa}, little Higgs models \cite{ArkaniHamed:2002qy, Schmaltz:2005ky}, grand unified theories \cite{Hewett:1988xc}, and extra dimension models \cite{Contino:2006nn}. From the viewpoint of model building, one attractive reason is that the VLQs can avoid the problem of quantum anomaly. In contrast, the LQs are well motivated in the grand unified theories \cite{Pati:1974yy, Georgi:1974sy, Fritzsch:1974nn}. Meanwhile, the LQs can be the solution to the $(g-2)_{\mu}$, $B$ physics, and other flavour anomalies \cite{Dorsner:2016wpm, London:2021lfn}. If we consider the VLQ and scalar LQ simultaneously, there can be two sources of $W$ mass corrections. Furthermore, it is also possible to explain the $(g-2)_{\mu}$ at the same time. Here, we study the triplet LQ and triplet VLQ extended model.

In this paper, we build the model in Sec.~\ref{sec:model} firstly. In Sec.~\ref{sec:contributions}, we calculate the new physics contributions to the $W$ boson mass and $(g-2)_{\mu}$. Then, we perform the numerical analysis in Sec.~\ref{sec:numerical}. Finally, we give the summary and conclusions in Sec.~\ref{sec:summary}.

\section{Model setup}\label{sec:model}
The SM gauge group is $SU_C(3)\otimes SU_L(2)\otimes U_Y(1)$, then new particles can carry different representations under this group. There are typically six types of scalar LQs \cite{Dorsner:2016wpm, Crivellin:2021ejk} and seven types of VLQs \cite{Aguilar-Saavedra:2013qpa}. In our previous paper \cite{He:2021yck}, we considered the $S_3$ LQ and $(X,T,B)_{L,R}$ VLQ extended model to explain the $(g-2)_{\mu}$ ahead of the CDF $W$ mass anomaly. Here, we investigate this model again, which is named as the $S_3+(X,T,B)_{L,R}$ for convenience.

The representation of $S_3$ is $(\bar{3},3,1/3)$ and it is $(3,3,2/3)$ for the $(X,T,B)_{L,R}$ \footnote{The singlet VLQ $T_{L,R}$ has been considered in the  Refs. \cite{Gu:2022htv, Balkin:2022glu, Cao:2022mif}.}. Then, the relevant Lagrangian can be decomposed as $\mathcal{L}_H^{Yukawa}+\mathcal{L}_{S_3}^{Yukawa}+\mathcal{L}_{XTB}^{gauge}+\mathcal{L}_{S_3}^{gauge}+\mathcal{L}_{S_3}^{scalar}$. Here, $\mathcal{L}_H^{Yukawa}$, $\mathcal{L}_{S_3}^{Yukawa}$, $\mathcal{L}_{XTB}^{gauge}$, $\mathcal{L}_{S_3}^{gauge}$, and $\mathcal{L}_{S_3}^{scalar}$ mark the VLQ Yukawa interactions with Higgs, VLQ Yukawa interactions with LQ, VLQ gauge interactions, LQ gauge interactions, and LQ scalar sector interactions, respectively. Below, we study these interactions carefully.
\subsection{VLQ Yukawa interactions with Higgs}\label{sec:VLQ:Higgs}
Firstly, let us write down the Yukawa interactions with Higgs.
\begin{align}
\mathcal{L}\supset -M_T\overline{(X,T,B)_L}\left(\begin{array}{c}X\\T\\B\end{array}\right)_{R}-y_{ij}^u\overline{Q_L}^iu_R^j\widetilde{\phi}-y_{ij}^d\overline{Q_L}^id_R^j\phi-y_{iT}\overline{(Q_L)}^i\Psi_R\widetilde{\phi}+\mathrm{h.c.}.
\end{align}
Here, we define $\widetilde{\phi}\equiv i\sigma^2\phi^{\ast}$, and $\sigma^a(a=1,2,3)$ are the Pauli matrices. The SM Higgs doublet $\phi$ is parameterized as $\phi=[0,(v+h)/{\sqrt{2}}]^T$ in the unitary gauge. The $Q_L^i,u_R^i,d_R^i$ represent the SM quark fields. The triplet $(X,T,B)_{L,R}$ can be parameterized in the following form:
\begin{align}
\Psi_{L,R}\equiv\Big(\begin{array}{cc}T_{L,R}&\sqrt{2}X_{L,R}\\ \sqrt{2}B_{L,R}&-T_{L,R}\end{array}\Big).
\end{align}
For simplicity, we only consider the mixing between third generation and VLQ \cite{Aguilar-Saavedra:2013qpa, He:2020suf, He:2020fqj}. After the electroweak symmetry breaking (EWSB), we obtain the following mass terms:
\begin{align}\label{eqn:quark:mass}
\mathcal{L}_{mass}\supset-
\left[\begin{array}{cc} \bar{t}_L&\bar{T}_L \end{array}\right]
\left[\begin{array}{cc}\frac{1}{\sqrt{2}}y^u_{33}v&\frac{1}{\sqrt{2}}y_{3T}v\\ 0 &M_T\end{array}\right]
\left[\begin{array}{cc} t_R\\T_R \end{array}\right]-
\left[\begin{array}{cc} \bar{b}_L&\bar{B}_L \end{array}\right]
\left[\begin{array}{cc}\frac{1}{\sqrt{2}}y^d_{33}v&y_{3T}v\\ 0 &M_T\end{array}\right]
\left[\begin{array}{cc} b_R\\B_R \end{array}\right]+\mathrm{h.c.}~.
\end{align}
Then, we can rotate the quark fields into mass eigenstates through the following transformations:
\begin{align}\label{eqn:tTquark:rotation}
\left[\begin{array}{c}t_L\\T_L\end{array}\right]\rightarrow
	\left[\begin{array}{cc}\cos\theta_L^t&\sin\theta_L^t\\-\sin\theta_L^t&\cos\theta_L^t\end{array}\right]
	\left[\begin{array}{c}t_L\\T_L\end{array}\right],\quad
\left[\begin{array}{c}t_R\\T_R\end{array}\right]\rightarrow
	\left[\begin{array}{cc}\cos\theta_R^t&\sin\theta_R^t\\-\sin\theta_R^t&\cos\theta_R^t\end{array}\right]
	\left[\begin{array}{c}t_R\\T_R\end{array}\right],
\end{align}
and
\begin{align}\label{eqn:bBquark:rotation}
\left[\begin{array}{c}b_L\\B_L\end{array}\right]\rightarrow
	\left[\begin{array}{cc}\cos\theta_L^b&\sin\theta_L^b\\-\sin\theta_L^b&\cos\theta_L^b\end{array}\right]
	\left[\begin{array}{c}b_L\\B_L\end{array}\right],\quad
\left[\begin{array}{c}b_R\\B_R\end{array}\right]\rightarrow
	\left[\begin{array}{cc}\cos\theta_R^b&\sin\theta_R^b\\-\sin\theta_R^b&\cos\theta_R^b\end{array}\right]
	\left[\begin{array}{c}b_R\\B_R\end{array}\right].
\end{align}
After the above quark transformations, we have the following mass eigenstate Yukawa interactions with Higgs:
\begin{align}
&\mathcal{L}_H^{Yukawa}\supset-\frac{m_t}{v}(c_L^t)^2h\bar{t}t-\frac{m_T}{v}(s_L^t)^2h\bar{T}T-\frac{m_b}{v}(c_L^b)^2h\bar{b}b-\frac{m_B}{v}(s_L^b)^2h\bar{B}B-\frac{m_T}{v}s_L^tc_L^th(\bar{t}_LT_R+\bar{T}_Rt_L)
\nonumber\\
&-\frac{m_t}{v}s_L^tc_L^th(\bar{T}_Lt_R+\bar{t}_RT_L)-\frac{m_B}{v}s_L^bc_L^bh(\bar{b}_LB_R+\bar{B}_Rb_L)-\frac{m_b}{v}s_L^bc_L^bh(\bar{B}_Lb_R+\bar{b}_RB_L).
\end{align}
In the above, the physical masses are labelled as $m_{t,T,b,B}$. The $s_L^{t(b)},c_L^{t(b)},s_R^{t(b)},c_R^{t(b)}$ are shorthands for $\sin\theta_L^{t(b)},\cos\theta_L^{t(b)},\sin\theta_R^{t(b)},\cos\theta_R^{t(b)}$, respectively. Besides, we have the following relations:
\begin{align}\label{eqn:HYukawa:relations}
&\tan\theta_R^t=\frac{m_t}{m_T}\tan\theta_L^t,~M_T^2=m_T^2(c_L^t)^2+m_t^2(s_L^t)^2,\nonumber\\
&\tan\theta_R^b=\frac{m_b}{m_B}\tan\theta_L^b,~M_T^2=m_B^2(c_L^b)^2+m_b^2(s_L^b)^2,\nonumber\\
&\qquad\sin2\theta_L^b=\frac{\sqrt{2}(m_T^2-m_t^2)}{m_B^2-m_b^2}\sin2\theta_L^t.
\end{align}
Hence, there are two new independent input parameters $m_T$ and $\theta_L^t$ (also denoted as $\theta_L$ in the following), and the parameters $M_T,m_B,\theta_R^t,\theta_L^b,\theta_R^b$ can be determined from the above equations (see App. \ref{app:VLQ:para}). One interesting thing is that the mass of $X$ quark is $M_T$, which is less than $m_T$ and $m_B$.

\subsection{VLQ Yukawa interactions with LQ}\label{sec:VLQ:LQ}
Now, let us consider the Yukawa interactions with LQ. In this $S_3+(X,T,B)_{L,R}$ model, the gauge eigenstate interactions can be written as
\begin{align}
\mc{L}\supset x_{ij}\overline{(Q_L)^C}^{i,a}(i\sigma^2)^{ab}(S_3)^{bc}L_L^{j,c}+x_{Ti}\mr{Tr}[\overline{(\Psi_R)^C}S_3]e_R^i+\mathrm{h.c.}.
\end{align}
Here, the $L_L^i,e_R^i$ stand for the SM lepton fields. Similarly, we also parameterize the $S_3$ triplet in the following form:
\begin{align}
S_3\equiv\Big(\begin{array}{cc}S_3^{1/3}&\sqrt{2}S_3^{4/3}\\ \sqrt{2}S_3^{-2/3}&-S_3^{1/3}\end{array}\Big).
\end{align}
In the above, the matrix elements $\big(\overline{(\Psi_R)^C}\big)_{ij}$ are defined as $\overline{\big((\Psi_R)_{ij}\big)^C}$. After the EWSB, the related Lagrangian can be reparameterized as
\begin{align}
&\mc{L}\supset y_L^{S_3\mu T}\bar{\mu}~\omega_-~T^C(S_3^{1/3})^\ast+y_R^{S_3\mu t}\bar{\mu}~\omega_+~t^C(S_3^{1/3})^\ast+y_L^{S_3\mu T}\bar{\mu}~\omega_-~B^C(S_3^{4/3})^\ast+\sqrt{2}y_R^{S_3\mu t}\bar{\mu}~\omega_+~b^C(S_3^{4/3})^\ast\nonumber\\
&+y_L^{S_3\mu T}\bar{\mu}~\omega_-~X^C(S_3^{-2/3})^\ast-\sqrt{2}y_R^{S_3\mu t}\overline{(\nu_{\mu})_L}~\omega_+~t^C(S_3^{-2/3})^\ast+y_R^{S_3\mu t}\overline{(\nu_{\mu})_L}~\omega_+~b^C(S_3^{1/3})^\ast+\mathrm{h.c.}.
\end{align}
Here, $\omega_\pm$ are the chirality operators $(1\pm\gamma^5)/2$. The new parameters $y_L^{S_3\mu T}$ and $y_R^{S_3\mu t}$ can be determined from the original $x_{ij}$ and $x_{Ti}$. We adopt the new parameters for convenience. When performing the transformations in Eqs. \eqref{eqn:tTquark:rotation} and  \eqref{eqn:bBquark:rotation}, we have the following mass eigenstate Yukawa interactions with LQ:
\begin{align}
&\mc{L}_{S^3}^{Yukawa}\supset\bar{\mu}(-y_L^{S_3\mu T}s_R^t\omega_-+y_R^{S_3\mu t}c_L^t\omega_+)t^C(S_3^{1/3})^\ast+\bar{\mu}(y_L^{S_3\mu T}c_R^t\omega_-+y_R^{S_3\mu t}s_L^t\omega_+)T^C(S_3^{1/3})^\ast\nonumber\\
&+\bar{\mu}(-y_L^{S_3\mu T}s_R^b\omega_-+\sqrt{2}y_R^{S_3\mu t}c_L^b\omega_+)b^C(S_3^{4/3})^\ast+\bar{\mu}(y_L^{S_3\mu T}c_R^b\omega_-+\sqrt{2}y_R^{S_3\mu t}s_L^b\omega_+)B^C(S_3^{4/3})^\ast\nonumber\\
&-\sqrt{2}y_R^{S_3\mu t}\overline{(\nu_{\mu})_L}~\omega_+(c_L^tt^C+s_L^tT^C)(S_3^{-2/3})^\ast+y_R^{S_3\mu t}\overline{(\nu_{\mu})_L}~\omega_+~(c_L^bb^C+s_L^bB^C)(S_3^{1/3})^\ast\nonumber\\
&+y_L^{S_3\mu T}\bar{\mu}~\omega_-~X^C(S_3^{-2/3})^\ast+\mathrm{h.c.}.
\end{align}

\subsection{VLQ gauge interactions}\label{sec:VLQ:gauge}
For the tripet VLQ $\Psi$, the covariant derivative of electroweak part is defined as $D_{\mu}\Psi=\partial_{\mu}\Psi-ig[W_{\mu}^a\tau^a,\Psi]-ig^{\prime}YB_{\mu}\Psi$, with $Y$ to be the $U_Y(1)$ charge and $\tau^a=\sigma^a/2$. The $W_{\mu}^a$ and $B_{\mu}$ label the $SU(2)_L$ and $U_Y(1)$ gauge fields, respectively. The gauge interactions are written as $\mr{Tr}(\bar{\Psi}i\slash\!\!\!\!D\Psi)/2$, in which the factor $1/2$ is to normalize the kinetic terms. After the EWSB, the charged current interactions can be written as
\begin{align}
\mc{L}\supset gW_{\mu}^+(\overline{T_L}\gamma^{\mu}B_L+\overline{T_R}\gamma^{\mu}B_R-\overline{X_L}\gamma^{\mu}T_L-\overline{X_R}\gamma^{\mu}T_R)+\mathrm{h.c.}.
\end{align}
Here, $W_{\mu}^\pm$ is defined as $(W_{\mu}^1\mp iW_{\mu}^2)/\sqrt{2}$. For the neutral current interactions, we perform the rotations $W_{\mu}^3=\cos\theta_W Z_{\mu}+\sin\theta_W A_{\mu}$ and $B_{\mu}=\cos\theta_W A_{\mu}-\sin\theta_W Z_{\mu}$, in which $\theta_W$ is the Weinberg angle. For convenience, the $\sin\theta_W(\cos\theta_W)$ is abbreviated as $s_W(c_W)$. Thus, the neutral current interactions can be written as
\begin{align}
&\mc{L}\supset\frac{g}{c_W}(-\frac{2}{3}s_W^2)Z_{\mu}(\overline{T_L}\gamma^{\mu}T_L+\overline{T_R}\gamma^{\mu}T_R)+\frac{g}{c_W}(-1+\frac{1}{3}s_W^2)Z_{\mu}(\overline{B_L}\gamma^{\mu}B_L+\overline{B_R}\gamma^{\mu}B_R)\nonumber\\
&+\frac{g}{c_W}(1-\frac{5}{3}s_W^2)(\overline{X_L}\gamma^{\mu}X_L+\overline{X_R}\gamma^{\mu}X_R).
\end{align}
As we know, the third generation quarks interact with $W$ and $Z$ bosons in the following form:
\begin{align}
&\mc{L}\supset \frac{g}{\sqrt{2}}W_{\mu}^+(\overline{t_L}\gamma^{\mu}b_L+\mathrm{h.c.})+\frac{g}{c_W}Z_{\mu}[(\frac{1}{2}-\frac{2}{3}s_W^2)\overline{t_L}\gamma^{\mu}t_L-\frac{2}{3}s_W^2\overline{t_R}\gamma^{\mu}t_R]\nonumber\\
&+\frac{g}{c_W}Z_{\mu}[(-\frac{1}{2}+\frac{1}{3}s_W^2)\overline{b_L}\gamma^{\mu}b_L+\frac{1}{3}s_W^2\overline{b_R}\gamma^{\mu}b_R].
\end{align}
After rotating the quark fields with Eqs. \eqref{eqn:tTquark:rotation} and  \eqref{eqn:bBquark:rotation}, we have the following mass eigenstate charged current interactions \footnote{The $WXt$ and $WXT$ interactions show sign difference from those in Ref. \cite{Aguilar-Saavedra:2013qpa}, while it can be absorbed through the redefinition of $X$ field and has no physical effects.}:
\begin{align}
&\mc{L}_{XTB}^{gauge}\supset\frac{g}{\sqrt{2}}W_{\mu}^+\Big\{\bar{t}\gamma^{\mu}[(c_L^tc_L^b+\sqrt{2}s_L^ts_L^b)\omega_-+\sqrt{2}s_R^ts_R^b\omega_+]b+\bar{t}\gamma^{\mu}[(c_L^ts_L^b-\sqrt{2}s_L^tc_L^b)\omega_--\sqrt{2}s_R^tc_R^b\omega_+]B\nonumber\\
&+\bar{T}\gamma^{\mu}[(s_L^tc_L^b-\sqrt{2}c_L^ts_L^b)\omega_--\sqrt{2}c_R^ts_R^b\omega_+]b+\bar{T}\gamma^{\mu}[(s_L^ts_L^b+\sqrt{2}c_L^tc_L^b)\omega_-+\sqrt{2}c_R^tc_R^b\omega_+]B\Big\}\nonumber\\
&+gW_{\mu}^+[\overline{X_L}\gamma^{\mu}(s_L^tt_L-c_L^tT_L)+\overline{X_R}\gamma^{\mu}(s_R^tt_R-c_R^tT_R)]+\mathrm{h.c.}.
\end{align}
Similarly, we have the following mass eigenstate neutral current interactions:
\begin{align}
&\mc{L}_{XTB}^{gauge}\supset\frac{g}{2c_W}Z_{\mu}\Big\{\bar{t}\gamma^{\mu}[\big((c_L^t)^2-\frac{4}{3}s_W^2\big)\omega_--\frac{4}{3}s_W^2\omega_+]t+\bar{T}\gamma^{\mu}[\big((s_L^t)^2-\frac{4}{3}s_W^2\big)\omega_--\frac{4}{3}s_W^2\omega_+]T\nonumber\\
&+s_L^tc_L^t(\overline{t_L}\gamma^{\mu}T_L+\overline{T_L}\gamma^{\mu}t_L)]+\bar{b}\gamma^{\mu}[\big(-1-(s_L^b)^2+\frac{2}{3}s_W^2\big)\omega_-+\big(-2(s_R^b)^2+\frac{2}{3}s_W^2\big)\omega_+]b\nonumber\\
&+\bar{B}\gamma^{\mu}[(-1-(c_L^b)^2+\frac{2}{3}s_W^2)\omega_-+\big(-2(c_R^b)^2+\frac{2}{3}s_W^2\big)\omega_+]B+s_L^bc_L^b(\overline{b_L}\gamma^{\mu}B_L+\overline{B_L}\gamma^{\mu}b_L)\nonumber\\
&+2s_R^bc_R^b(\overline{b_R}\gamma^{\mu}B_R+\overline{B_R}\gamma^{\mu}b_R)+2(1-\frac{5}{3}s_W^2)(\overline{X_L}\gamma^{\mu}X_L+\overline{X_R}\gamma^{\mu}X_R)\Big\}.
\end{align}
\subsection{LQ gauge interactions}\label{sec:LQ:gauge}
For the LQ $S_3$, the covariant derivative of electroweak part is defined as
$D_{\mu}S_3=\partial_{\mu}S_3-ig[W_{\mu}^a\tau^a,S_3]-ig^{\prime}YB_{\mu}S_3$. Then, the gauge interactions $\mathcal{L}_{S_3}^{gauge}\supset\frac{1}{2}\mr{Tr}[(D_{\mu}S_3)^{\dag}(D^{\mu}S_3)]$ can be expanded as follows.
\begin{itemize}[itemindent=-10pt]
\item $S_3S_3W$ interaction:
\begin{align}
igW_{\mu}^+[(\partial^{\mu}S_3^{4/3})^{\ast}S_3^{1/3}-(\partial^{\mu}S_3^{1/3})(S_3^{4/3})^{\ast}+(\partial^{\mu}S_3^{-2/3})(S_3^{1/3})^{\ast}-(\partial^{\mu}S_3^{1/3})^{\ast}S_3^{-2/3}]+\mathrm{h.c.}.
\end{align}
\item $S_3S_3Z$ interaction:
\begin{align}
&\frac{igZ_{\mu}}{c_W}\Big\{-\frac{1}{3}s_W^2[(\partial^{\mu}S_3^{1/3})(S_3^{1/3})^{\ast}-(\partial^{\mu}S_3^{1/3})^{\ast}S_3^{1/3}]+(\frac{2}{3}s_W^2-1)[(\partial^{\mu}S_3^{-2/3})(S_3^{-2/3})^{\ast}-(\partial^{\mu}S_3^{-2/3})^{\ast}S_3^{-2/3}]\nonumber\\
&+(1-\frac{4}{3}s_W^2)[(\partial^{\mu}S_3^{4/3})(S_3^{4/3})^{\ast}-(\partial^{\mu}S_3^{4/3})^{\ast}S_3^{4/3}]\Big\}.
\end{align}
\item $S_3S_3\gamma$ interaction:
\begin{align}
&ieA_{\mu}\Big\{\frac{1}{3}[(\partial^{\mu}S_3^{1/3})(S_3^{1/3})^{\ast}-(\partial^{\mu}S_3^{1/3})^{\ast}S_3^{1/3}]-\frac{2}{3}[(\partial^{\mu}S_3^{-2/3})(S_3^{-2/3})^{\ast}-(\partial^{\mu}S_3^{-2/3})^{\ast}S_3^{-2/3}]\nonumber\\
&+\frac{4}{3}[(\partial^{\mu}S_3^{4/3})(S_3^{4/3})^{\ast}-(\partial^{\mu}S_3^{4/3})^{\ast}S_3^{4/3}]\Big\}.
\end{align}
\item $S_3S_3WW$ interaction:
\begin{align}
&g^2\Big\{W_{\mu}^+W^{-,{\mu}}[2S_3^{1/3}(S_3^{1/3})^{\ast}+S_3^{-2/3}(S_3^{-2/3})^{\ast}+S_3^{4/3}(S_3^{4/3})^{\ast}]\nonumber\\
&-S_3^{4/3}(S_3^{-2/3})^{\ast}W_{\mu}^-W^{-,{\mu}}-S_3^{-2/3}(S_3^{4/3})^{\ast}W_{\mu}^+W^{+,{\mu}}\Big\}.
\end{align}
\item $S_3S_3WZ$ interaction:
\begin{align}
&\frac{g^2}{c_W}W_{\mu}^+Z^{\mu}[(-1+\frac{1}{3}s_W^2)S_3^{-2/3}(S_3^{1/3})^{\ast}+(-1+\frac{5}{3}s_W^2)S_3^{1/3}(S_3^{4/3})^{\ast}]+\mathrm{h.c.}.
\end{align}
\item $S_3S_3W\gamma$ interaction:
\begin{align}
&egW_{\mu}^+A^{\mu}[-\frac{1}{3}S_3^{-2/3}(S_3^{1/3})^{\ast}-\frac{5}{3}S_3^{1/3}(S_3^{4/3})^{\ast}]+\mathrm{h.c.}.
\end{align}
\item $S_3S_3ZZ$ interaction:
\begin{align}
&\frac{g^2}{c_W^2}Z_{\mu}Z^{\mu}[\frac{1}{9}s_W^4S_3^{1/3}(S_3^{1/3})^{\ast}+(1-\frac{4}{3}s_W^2)^2S_3^{4/3}(S_3^{4/3})^{\ast}+(1-\frac{2}{3}s_W^2)^2S_3^{-2/3}(S_3^{-2/3})^{\ast}].
\end{align}
\item $S_3S_3Z\gamma$ interaction:
\begin{align}
&\frac{2eg}{c_W}Z_{\mu}A^{\mu}[-\frac{1}{9}s_W^2S_3^{1/3}(S_3^{1/3})^{\ast}+\frac{4}{3}(1-\frac{4}{3}s_W^2)S_3^{4/3}(S_3^{4/3})^{\ast}-\frac{2}{3}(-1+\frac{2}{3}s_W^2)S_3^{-2/3}(S_3^{-2/3})^{\ast}].
\end{align}
\item $S_3S_3\gamma\gamma$ interaction:
\begin{align}
&e^2A_{\mu}A^{\mu}[\frac{1}{9}S_3^{1/3}(S_3^{1/3})^{\ast}+\frac{16}{9}S_3^{4/3}(S_3^{4/3})^{\ast}+\frac{4}{9}S_3^{-2/3}(S_3^{-2/3})^{\ast}].
\end{align}
\end{itemize}
\subsection{LQ scalar sector interactions}\label{sec:LQ:scalar}
Here, we consider the scalar sector interactions. The mass related terms can be written as \footnote{There is another interaction term $\phi^{\dag}\mr{Tr}_C[S_3(S_3)^{\dag}]\phi$, in which the $\mr{Tr}_C$ only acts on the $SU_C(3)$ color space. However, it can be removed since we have the relation $(\phi^{\dag}\phi)\mr{Tr}[(S_3)^{\dag}S_3]=\phi^{\dag}(S_3)^{\dag}S_3\phi+\phi^{\dag}\mr{Tr}_C[S_3(S_3)^{\dag}]\phi$.}
\begin{align}
&\mathcal{L}_{S_3}^{scalar}\supset-\frac{1}{2}m_{S_3}^2\mr{Tr}[(S_3)^{\dag}S_3]-\lambda_{\phi S_3}(\phi^{\dag}\phi)\mr{Tr}[(S_3)^{\dag}S_3]-\widetilde{\lambda}_{\phi S_3}\phi^{\dag}(S_3)^{\dag}S_3\phi.
\end{align}
After the EWSB, we have the following mass equations \cite{Bandyopadhyay:2021kue}:
\begin{align}\label{eqn:scalar:mass}
&m_{S_3^{4/3}}^2=m_{S_3}^2+\lambda_{\phi S_3}v^2+\widetilde{\lambda}_{\phi S_3}v^2,\quad m_{S_3^{1/3}}^2=m_{S_3}^2+\lambda_{\phi S_3}v^2+\frac{1}{2}\widetilde{\lambda}_{\phi S_3}v^2,\quad m_{S_3^{-2/3}}^2=m_{S_3}^2+\lambda_{\phi S_3}v^2.
\end{align}
Obviously, the $(\phi^{\dag}\phi)\mr{Tr}[(S_3)^{\dag}S_3]$ term does not contribute to the mass splittings. While, there are tree level generated mass splittings, which are controlled by the coupling $\widetilde{\lambda}_{\phi S_3}$. In fact, this is similar to the traditional colorless elctroweak triplet. The difference is that the mass splittings for the traditional electroweak triplet can also be caused by the non-zero triplet vacuum expectation value \cite{Aoki:2011pz, Arhrib:2011uy}. In Ref. \cite{Athron:2022qpo}, the authors studied the mass splittings originating from the $S_1$ and $S_3$ LQ mixing.

Although there are three mass parameters $m_{S_3^{4/3}}^2$, $m_{S_3^{1/3}}^2$, and $m_{S_3^{-2/3}}^2$, only two of them are relevant because we can redefine the mass. For convenience, let us define the following mass splitting quantity:
\begin{align}\label{eqn:scalar:deltam}
\Delta m^2(\approx 2m_{S_3^{1/3}}\cdot\Delta m)\equiv m_{S_3^{4/3}}^2-m_{S_3^{1/3}}^2=m_{S_3^{1/3}}^2-m_{S_3^{-2/3}}^2=\frac{1}{2}\widetilde{\lambda}_{\phi S_3}v^2.
\end{align}
Thus, we can choose the input parameters to be $(m_{S_3^{1/3}},\widetilde{\lambda}_{\phi S_3})$ or $(m_{S_3^{1/3}},\Delta m)$.

\section{Explanation of the $W$ boson mass and $(g-2)_{\mu}$ anomalies}\label{sec:contributions}
\subsection{Contributions to the $W$ boson mass}
If we choose the $\{\alpha,G_F,m_Z\}$ scheme, the $W$ boson mass can be determined from the formula \cite{Awramik:2003rn, Erler:2019hds}:
\begin{align}
m_W^2=\frac{m_Z^2}{2}\Big(1+\sqrt{1-\frac{\sqrt{8}\pi\alpha(1+\Delta r)}{G_Fm_Z^2}}\Big),
\end{align}
Then, we have the following approximation:
\begin{align}
\frac{\Delta m_W^2}{m_W^2}=-\frac{s_W^2}{c_W^2-s_W^2}\delta r.
\end{align}
In the above, we define the quantities $\Delta m_W^2\equiv (m_W^{\mr{NP}})^2-(m_W^{\mr{SM}})^2$ and $\delta r\equiv\Delta r^{\mr{NP}}-\Delta r^{\mr{SM}}$ to isolate the new physics contributions. If we neglect the new physics contributions from wave function renormalization constants, vertex, and box diagrams in the $\mu$ decay \footnote{Strictly speaking, we need to analyse the complete new physics corrections to the $\mu$ decay. Here, we will not study that hard work.}, the $W$ mass correction can be correlated with $S,T,U$ oblique parameters as \cite{Peskin:1990zt, Peskin:1991sw, Maksymyk:1993zm, Burgess:1993mg, Grimus:2008nb, Heinemeyer:2013dia}
\begin{align}\label{eqn:mW:STU}
\frac{\Delta m_W^2}{m_W^2}=\frac{2\Delta m_W}{m_W}=\frac{\alpha}{c_W^2-s_W^2}(-\frac{1}{2}\Delta S+c_W^2\Delta T+\frac{c_W^2-s_W^2}{4s_W^2}\Delta U),
\end{align}
where we define the deviations $\Delta S\equiv S^{\mr{NP}}-S^{\mr{SM}}$, $\Delta T\equiv T^{\mr{NP}}-T^{\mr{SM}}$, and $\Delta U\equiv U^{\mr{NP}}-U^{\mr{SM}}$. In most cases, the $T$ parameter dominates. There are mainly two part contributions to the oblique parameters in the $S_3+(X,T,B)_{L,R}$ model. One is from the LQ contributions, and the other is from the VLQ.

Now, let us turn to the oblique contributions from the LQ loops, which are denoted as $\Delta S^{S_3}$, $\Delta T^{S_3}$, and $\Delta U^{S_3}$. Unlike the traditional electroweak Higgs triplet model, the $S_3$ does not modify the $T$ parameter at tree level because of the exact color symmetry. The complete one-loop results can be calculated through the interactions given in Sec. \ref{sec:LQ:gauge}, and the details are given in App. \ref{app:LQ:ST}. The $U$ parameter formula is lengthy, and the $S$ and $T$ parameters have the following compact expressions:
\begin{align}\label{eqn:mW:STU:S3}
\Delta S^{S_3}=-\frac{N_C}{9\pi}\log\frac{m_{S_3^{4/3}}^2}{m_{S_3^{-2/3}}^2},\qquad\Delta T^{S_3}=\frac{N_C}{8\pi m_W^2s_W^2}[\theta_+(m_{S_3^{4/3}}^2,m_{S_3^{1/3}}^2)+\theta_+(m_{S_3^{-2/3}}^2,m_{S_3^{1/3}}^2)].
\end{align}
Here, $N_C=3$ is the color factor, and the function $\theta_+$ is defined as
\begin{align}
\theta_+(y_1,y_2)\equiv y_1+y_2-\frac{2y_1y_2}{y_1-y_2}\log\frac{y_1}{y_2}.
\end{align}
Obviously, we have $\Delta T^{S_3}\ge0$ because of the inequality $\theta_+(x,y)\ge0$, in which the equality applies if and only if $x=y$. For the $\Delta S^{S_3}$, it is negative if $m_{S_3^{4/3}}>m_{S_3^{-2/3}}$ ($\widetilde{\lambda}_{\phi S_3}>0$) and positive if $m_{S_3^{4/3}}<m_{S_3^{-2/3}}$ ($\widetilde{\lambda}_{\phi S_3}<0$). The mass expressions are shown in Eqs. \eqref{eqn:scalar:mass} and \eqref{eqn:scalar:deltam}. In the approximation of $\widetilde{\lambda}_{\phi S_3}v^2\ll m_{S_3^{1/3}}^2$ (or $\Delta m\ll m_{S_3^{1/3}}$), the $S$, $T$, and $U$ parameters can be expanded as
\begin{align}\label{eqn:mW:STU:S3app}
&\qquad\qquad\qquad\Delta S^{S_3}\approx-\frac{\widetilde{\lambda}_{\phi S_3}v^2}{3\pi m_{S_3^{1/3}}^2}\approx-\frac{4\Delta m}{3\pi m_{S_3^{1/3}}},\nonumber\\
&\Delta T^{S_3}\approx\frac{(\widetilde{\lambda}_{\phi S_3})^2v^4}{16\pi s_W^2m_W^2m_{S_3^{1/3}}^2}\approx\frac{(\Delta m)^2}{\pi s_W^2m_W^2},\qquad\Delta U^{S_3}\approx\frac{7(\widetilde{\lambda}_{\phi S_3})^2v^4}{40\pi m_{S_3^{1/3}}^4}\approx\frac{14(\Delta m)^2}{5\pi m_{S_3^{1/3}}^2}.
\end{align}
In the limit $\widetilde{\lambda}_{\phi S_3}\rightarrow0$, they vanish. These results completely agree with those in Refs. \cite{Dorsner:2019itg, Crivellin:2020ukd}.

For the VLQ part, the oblique corrections are caused by both the modification of SM quark gauge couplings and new VLQ loops, which are denoted as $\Delta S^{XTB}$, $\Delta T^{XTB}$, and $\Delta U^{XTB}$. Their analytic expressions are lengthy; hence, the details are given in App. \ref{app:VLQ:ST}. Considering $m_b\ll m_t\ll m_T$ and $s_L^t\ll1$, the formulae can be approximated as
\begin{align}\label{eqn:XTB:STapp}
&\qquad\qquad\qquad\Delta S^{XTB}\approx\frac{N_C(s_L^t)^2}{18\pi}(-12\log\frac{m_T}{m_t}-16\log\frac{m_t}{m_b}+29),\nonumber\\
&\Delta T^{XTB}\approx\frac{N_Cm_t^2(s_L^t)^2}{8\pi s_W^2m_W^2}(6\log\frac{m_T}{m_t}-5),\qquad\Delta U^{XTB}\approx\frac{N_C(s_L^t)^2}{18\pi}(24\log\frac{m_t}{m_b}-5).
\end{align}
If $s_L^t$ goes to zero, the $\Delta S^{XTB}$, $\Delta T^{XTB}$, and $\Delta U^{XTB}$ will vanish. Our expansion results agree with those in Cao's paper \cite{Cao:2022mif}, while the expansion of $S$ parameter differs from the result in Ref. \cite{Chen:2017hak}.\footnote{In Ref. \cite{Chen:2017hak}, the authors adopt the $S$, $T$, and $U$ parameter formulae given in Ref. \cite{Lavoura:1992np} directly. For the $(X,T,B)_{L,R}$ triplet, the $T$ parameter formula still holds, while the $S$ and $U$ parameter formulae are no longer applicable. The detailed clarification is given in App. \ref{app:VLQ:ST}}. According to Eq. \eqref{eqn:XTB:STapp}, the $\Delta T^{XTB}$ and $\Delta U^{XTB}$ are always positive, while the $\Delta S^{XTB}$ is always negative. Typically, the $\Delta T^{XTB}$ is several times larger than $\Delta S^{XTB}$ and $\Delta U^{XTB}$ because of the $m_t^2/(m_W^2s_W^2)$ factor.

After summing the fermion and boson contributions, we can obtain the total oblique parameter deviations as $\Delta S\equiv\Delta S^{XTB}+\Delta S^{S_3}$, $\Delta T\equiv\Delta T^{XTB}+\Delta T^{S_3}$, and $\Delta U\equiv\Delta U^{XTB}+\Delta U^{S_3}$. To explain the $W$ boson mass anomaly, a positive $\Delta T$ is required, which is satisfied for both the VLQ and LQ contributions. There are four independent parameters involved in the oblique corrections: $m_T$ and $s_L$ for the VLQ, and $m_{S_3^{1/3}}$ and $\widetilde{\lambda}_{\phi S_3}$ for the LQ.
 
\subsection{Contributions to the $(g-2)_{\mu}$}
According to the BNL and FNAL experiments \cite{Muong-2:2006rrc, Muong-2:2021ojo}, the most recent muon anomalous magnetic dipole moment is measured as $a_{\mu}^{\mr{Exp}}=116592061(41)\times10^{-11}$. In the SM, it is predicted as $a_{\mu}^{\mr{SM}}=116591810(43)\times10^{-11}$ \cite{Aoyama:2020ynm}. Thus, the deviation is $\Delta a_{\mu}\equiv a_{\mu}^{\mr{Exp}}-a_{\mu}^{\mr{SM}}=(251\pm59)\times10^{-11}$, which means the $4.2\sigma$ discrepancy. LQ models can be the solution to this $(g-2)_{\mu}$ anomaly \cite{Chakraverty:2001yg, Cheung:2001ip, Djouadi:1989md, Dorsner:2016wpm, ColuccioLeskow:2016dox, Dorsner:2019itg, Dorsner:2020aaz, Crivellin:2020tsz}. In our previous paper \cite{He:2021yck}, we studied the $(g-2)_{\mu}$ in the $S_3+(X,T,B)_{L,R}$ model, in which the contributions are mainly from $T$ and $B$ quarks. Considering all the contributions from $t,T,b,B,X$ quarks, the complete expression is calculated as
\begin{align}
&\Delta a_{\mu}=\frac{m_{\mu}^2}{8\pi^2}\Bigg\{\frac{|y_L^{S_3\mu T}|^2(s_R^t)^2+|y_R^{S_3\mu t}|^2(c_L^t)^2}{m_{S_3^{1/3}}^2}f_{LL}^{S_3}(\frac{m_t^2}{m_{S_3^{1/3}}^2})-\frac{2m_t}{m_{\mu}}\frac{c_L^ts_R^t}{m_{S_3^{1/3}}^2}\mr{Re}[y_L^{S_3\mu T}(y_R^{S_3\mu t})^\ast]f_{LR}^{S_3}(\frac{m_t^2}{m_{S_3^{1/3}}^2})\nonumber\\
&+\frac{|y_L^{S_3\mu T}|^2(c_R^t)^2+|y_R^{S_3\mu t}|^2(s_L^t)^2}{m_{S_3^{1/3}}^2}f_{LL}^{S_3}(\frac{m_T^2}{m_{S_3^{1/3}}^2})+\frac{2m_T}{m_{\mu}}\frac{s_L^tc_R^t}{m_{S_3^{1/3}}^2}\mr{Re}[y_L^{S_3\mu T}(y_R^{S_3\mu t})^\ast]f_{LR}^{S_3}(\frac{m_T^2}{m_{S_3^{1/3}}^2})\nonumber\\
&+\frac{|y_L^{S_3\mu T}|^2(s_R^b)^2+2|y_R^{S_3\mu t}|^2(c_L^b)^2}{m_{S_3^{4/3}}^2}\widetilde{f}_{LL}^{S_3}(\frac{m_b^2}{m_{S_3^{4/3}}^2})-\frac{2\sqrt{2}m_b}{m_{\mu}}\frac{c_L^bs_R^b}{m_{S_3^{4/3}}^2}\mr{Re}[y_L^{S_3\mu T}(y_R^{S_3\mu t})^\ast]\widetilde{f}_{LR}^{S_3}(\frac{m_b^2}{m_{S_3^{4/3}}^2})\nonumber\\
&+\frac{|y_L^{S_3\mu T}|^2(c_R^b)^2+2|y_R^{S_3\mu t}|^2(s_L^b)^2}{m_{S_3^{4/3}}^2}\widetilde{f}_{LL}^{S_3}(\frac{m_B^2}{m_{S_3^{4/3}}^2})+\frac{2\sqrt{2}m_B}{m_{\mu}}\frac{s_L^bc_R^b}{m_{S_3^{4/3}}^2}\mr{Re}[y_L^{S_3\mu T}(y_R^{S_3\mu t})^\ast]\widetilde{f}_{LR}^{S_3}(\frac{m_B^2}{m_{S_3^{4/3}}^2})\nonumber\\
&+\frac{|y_L^{S_3\mu T}|^2}{m_{S_3^{-2/3}}^2}\hat{f}_{LL}^{S_3}(\frac{m_X^2}{m_{S_3^{-2/3}}^2})\Bigg\}.
\end{align}
In the above, the subscripts ``$LL$" and ``$LR$" denote the contributions without and with chiral enhancements. The related functions are defined as
\begin{align}
&f_{LL}^{S_3}(x)\equiv\frac{1+4x-5x^2+2x(2+x)\log x}{4(1-x)^4},\qquad\qquad\quad f_{LR}^{S_3}(x)\equiv-\frac{7-8x+x^2+(4+2x)\log x}{4(1-x)^3},\nonumber\\
&\widetilde{f}_{LL}^{S_3}(x)\equiv-\frac{2-7x+2x^2+3x^3+2x(1-4x)\log x}{4(1-x)^4},\quad\widetilde{f}_{LR}^{S_3}(x)\equiv-\frac{1+4x-5x^2-(2-8x)\log x}{4(1-x)^3},\nonumber\\
&\qquad\qquad\qquad\qquad\qquad\hat{f}_{LL}^{S_3}(x)\equiv\frac{4+x-8x^2+3x^3+2x(5-2x)\log x}{4(1-x)^4}.
\end{align}
Considering $m_b\ll m_t\ll m_T\approx m_B$, $\theta_L^t\ll1$, and $m_{S_3^{4/3}}\approx m_{S_3^{1/3}}\approx m_{S_3^{-2/3}}\approx m_{S_3}$, it can be approximated as
\begin{align}
&\Delta a_{\mu}\approx\frac{m_{\mu}m_T}{4\pi^2m_{S_3}^2}[f_{LR}^{S_3}(\frac{m_T^2}{m_{S_3}^2})+2\widetilde{f}_{LR}^{S_3}(\frac{m_T^2}{m_{S_3}^2})+\frac{m_t^2}{m_T^2}(\frac{7}{4}+\log \frac{m_t^2}{m_{S_3}^2})]\cdot\mr{Re}[y_L^{S_3\mu T}(y_R^{S_3\mu t})^\ast]s_L.
\end{align}
As we can see, the contributions to $(g-2)_{\mu}$ are mainly determined by the parameters $m_T$, $s_L$, and $m_{S_3^{1/3}}$. Although the parameter $\widetilde{\lambda}_{\phi S_3}$ can also alter the correction, it is sub-dominated through the LQ mass differences.
\section{Numerical analysis}\label{sec:numerical}

In this paper, we choose the SM input parameters to be $m_Z=91.1876\mathrm{GeV},m_W=80.379\mathrm{GeV},m_{\mu}=105.66\mathrm{MeV},
m_t=172.5\mathrm{GeV},m_b=4.2\mathrm{GeV},\alpha=1/128$, and $c_W=m_W/m_Z$ \cite{ParticleDataGroup:2020ssz}. Besides, we define the $W$ mass deviation quantity $\Delta m_W^{\mr{exp}}\equiv m_W^{\mr{exp}}-m_W^{\mr{SM}}$. Here, $m_W^{\mr{exp}}$ is the CDF result $80,433.5\pm9.4\mr{MeV}$ \cite{CDF:2022hxs}, and $m_W^{\mr{SM}}$ is the SM prediction $80,357\pm6\mr{MeV}$ \cite{ParticleDataGroup:2020ssz}. Then, the $\Delta m_W^{\mr{exp}}$ is calculated to be $76.5\pm11.2\mr{MeV}$. For the VLQ mass, the direct search requires it to be above 1.4 TeV \cite{CMS:2019eqb, ATLAS:2018ziw}. For the LQ mass, the direct search also requires it to be above 1.5 TeV \cite{CMS:2018lab, ATLAS:2020xov}. We also consider the constraints from electro-weak precision observables. Because it is small for the correlation between the oblique corrections and $Zbb$ couplings \cite{Ciuchini:2013pca, deBlas:2016ojx}, we can treat them separately for simplicity. There is $b-B$ mixing induced tree level modification of the $Zbb$ coupling, which leads to the bound $s_L\lesssim0.05$ \cite{Aguilar-Saavedra:2013qpa, Chen:2017hak}. In Ref. \cite{deBlas:2022hdk}, the oblique parameters are updated as follows (standard average scenario):
\begin{align}
\Delta S^{\mr{fit}}=0.005,\quad\sigma_S=0.096,\quad\Delta T^{\mr{fit}}=0.040,\quad\sigma_T=0.120,\quad\Delta U^{\mr{fit}}=0.134,\quad\sigma_U=0.087,
\end{align}
with the correlation matrix
\begin{align}
\rho=\left[\begin{array}{ccc}
1.00 & 0.91 & -0.65 \\
0.91 & 1.00 & -0.88 \\
-0.65 & -0.88 & 1.00
\end{array}\right].
\end{align}
Then, we can define the $\chi^2$ quantity as
\begin{align}
\chi^2\equiv\sum\limits_{i,j=1,2,3}\frac{O_i-O_i^{\mr{fit}}}{\sigma_i}(\rho^{-1})_{ij}\frac{O_j-O_j^{\mr{fit}}}{\sigma_j},
\end{align}
where the indices $1,2,3$ label the $S,T,U$ parameters. Next, we perform the $\chi^2$ fit of the oblique parameters.

\begin{figure}[!h]
\begin{center}
\includegraphics[scale=0.4]{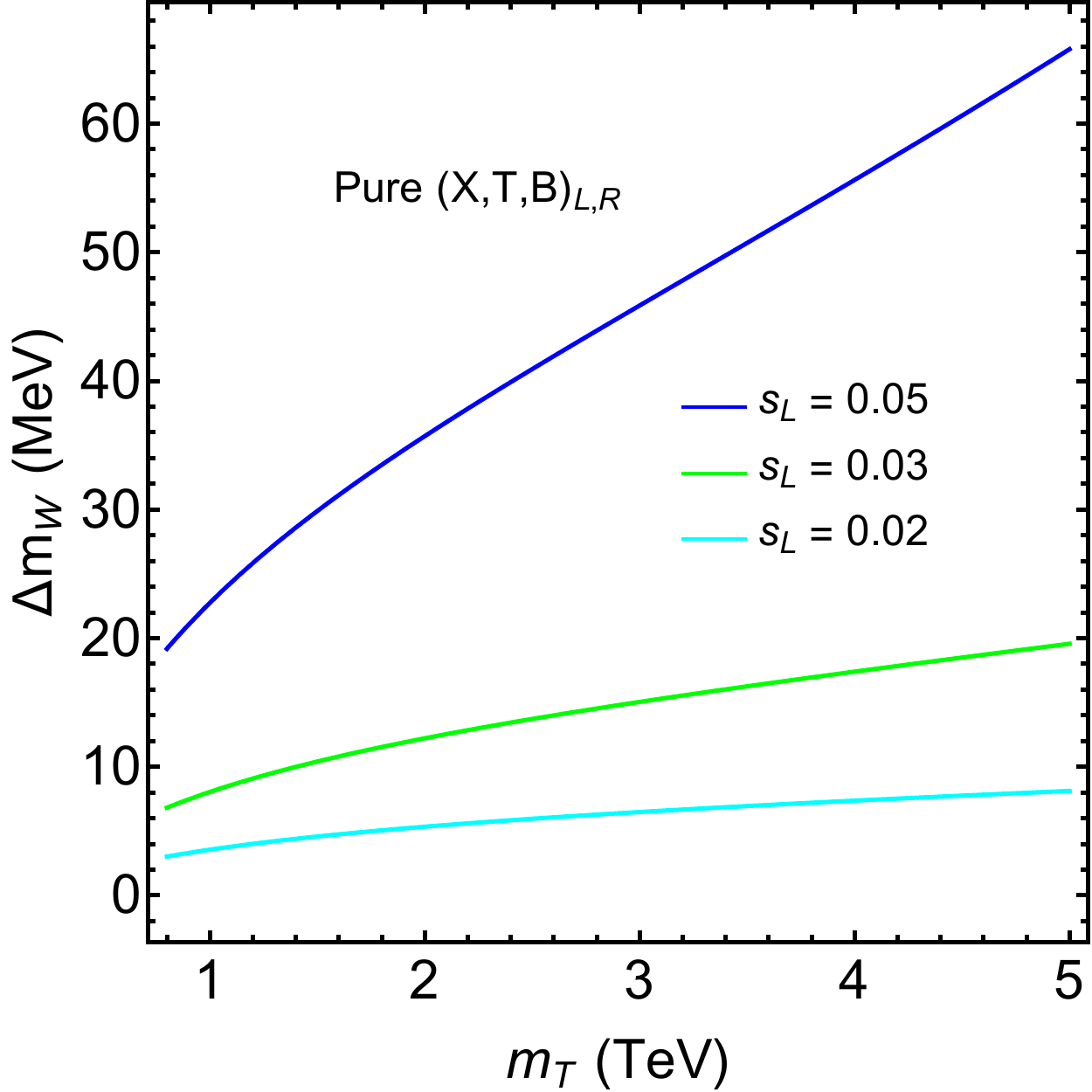}\qquad\qquad\qquad
\includegraphics[scale=0.4]{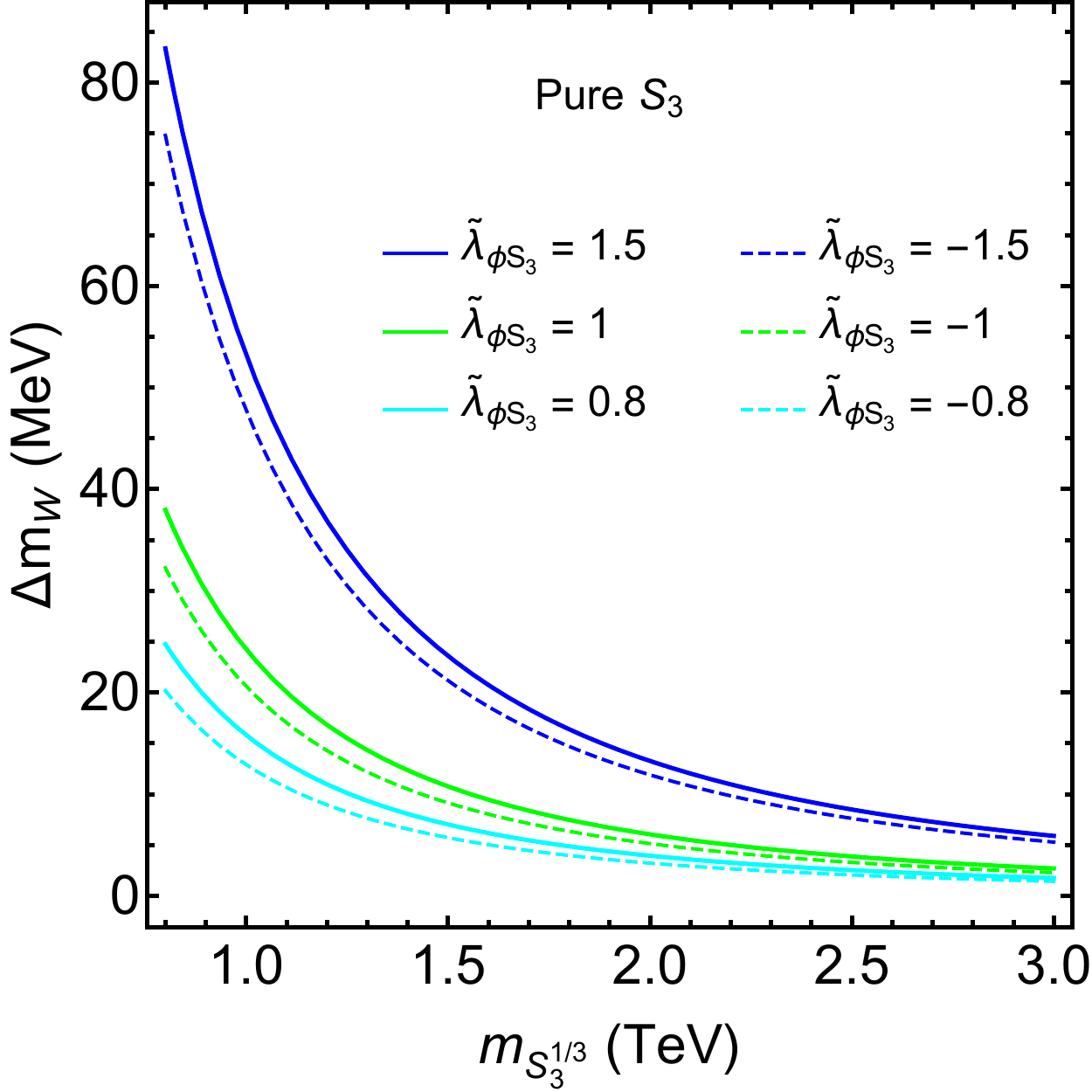}
\caption{The pure $(X,T,B)_{L,R}$ contributions to $\Delta m_W$ as a function of $m_T$ for different $s_L$ (left). The pure $S_3$ contributions to $\Delta m_W$ as a function of $m_{S_3^{1/3}}$ for different $\widetilde{\lambda}_{\phi S_3}$ (right).}\label{fig:mW:curve}
\end{center}
\end{figure}

\begin{figure}[!h]
\begin{center}
\includegraphics[scale=0.4]{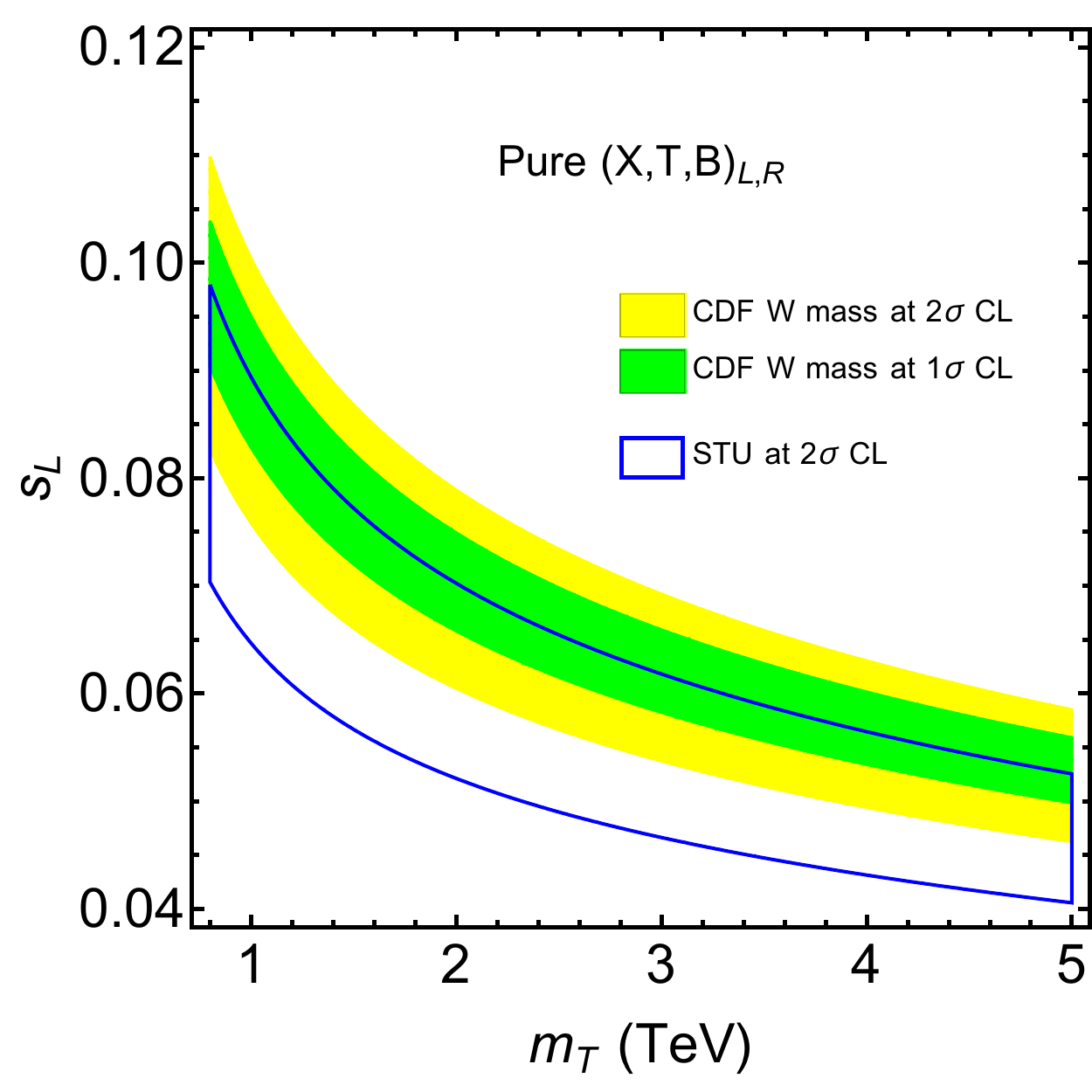}\\
\includegraphics[scale=0.4]{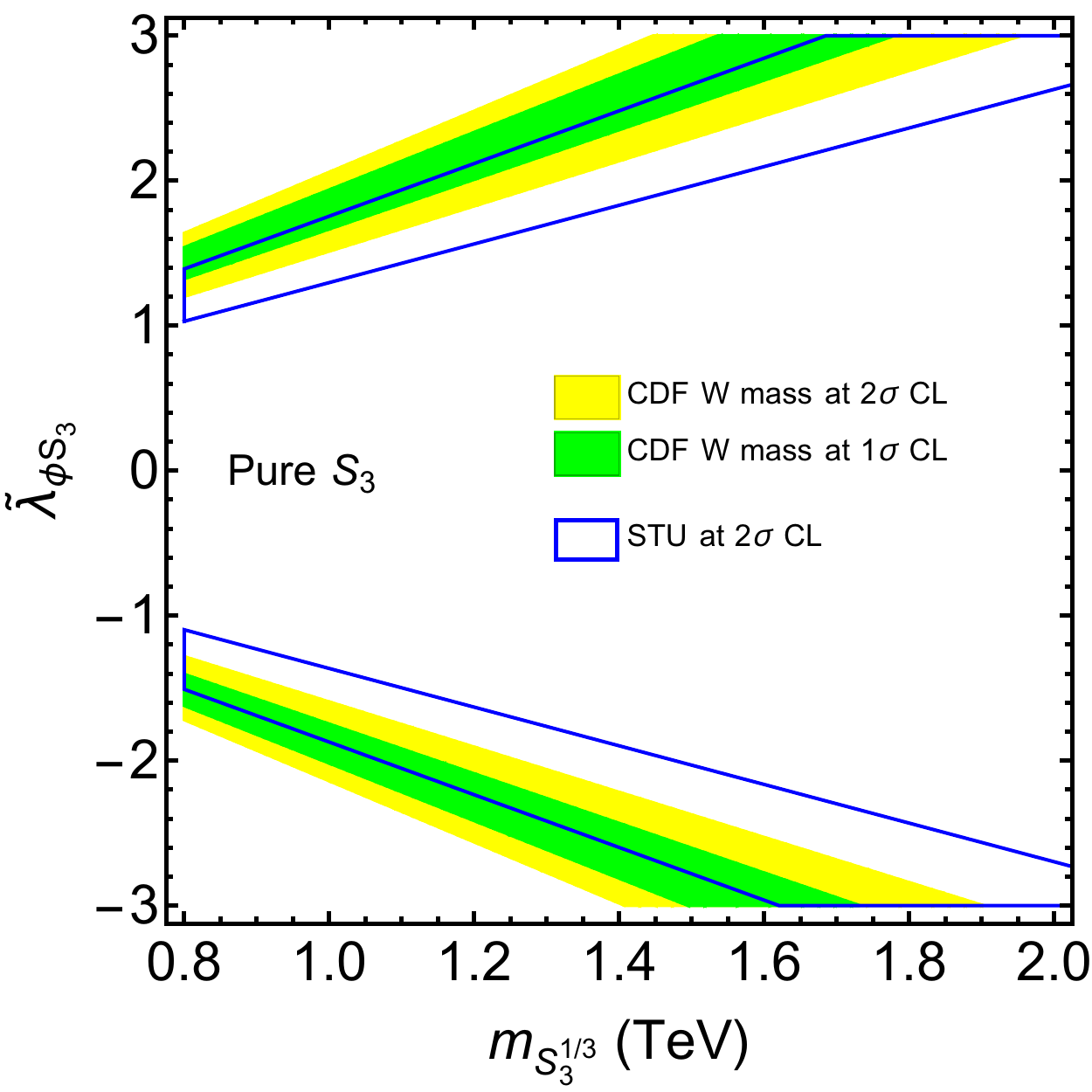}\qquad\qquad\qquad
\includegraphics[scale=0.4]{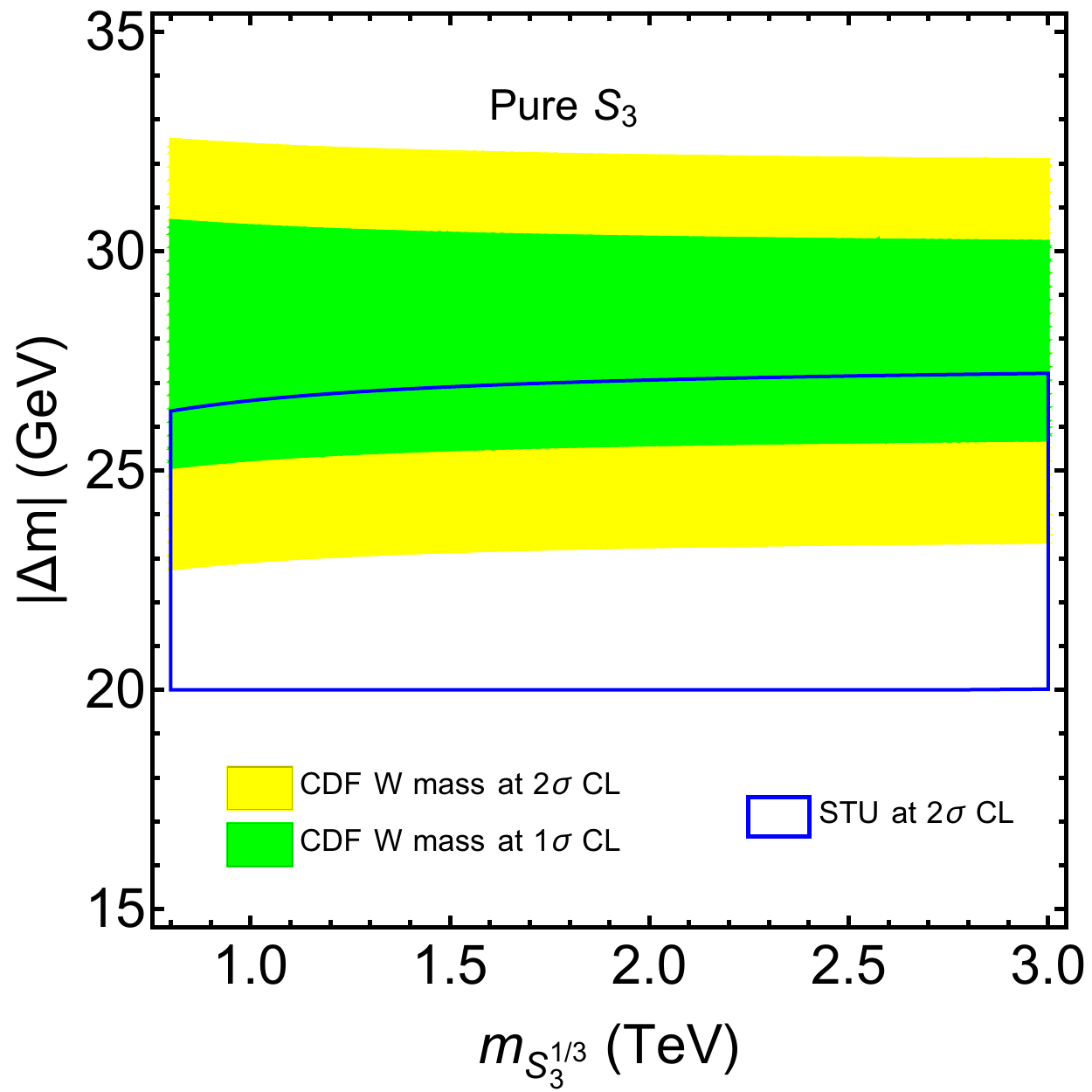}
\caption{The pure $(X,T,B)_{L,R}$ case in the plane of $m_T-s_L$ (upper), the pure $S_3$ case in the plane of $m_{S_3^{1/3}}-\widetilde{\lambda}_{\phi S_3}$ (lower left), and the pure $S_3$ case in the plane of $m_{S_3^{1/3}}-|\Delta m|$ (lower right). The CDF $W$ mass allowed parameter space is shown at $1\sigma$ (green) and $2\sigma$ (yellow) confidence levels (CLs), respectively. The blue line enclosed area is bounded by the $S,T,U$ parameters at $2\sigma$ CL.}\label{fig:mW:contour}
\end{center}
\end{figure}

First of all, let us compare the contributions from $(X,T,B)_{L,R}$ and $S_3$ roughly. In Fig. \ref{fig:mW:curve}, we show their individual contributions to the $W$ boson mass. For the pure $(X,T,B)_{L,R}$ case, the behaviour is as expected because we have the approximation $\Delta m_W\propto (s_L^t)^2\log(m_T/m_t)$. Thus, we need larger $s_L^t$ and $m_T$ to produce a sizable $W$ mass correction. For the pure $S_3$ case, the behaviour is also as expected because we have the approximation $\Delta m_W\propto (\widetilde{\lambda}_{\phi S_3})^2/m_{S_3^{1/3}}^2$ if $\widetilde{\lambda}_{\phi S_3}\sim\mc{O}(1)$. Thus, we need larger $\widetilde{\lambda}_{\phi S_3}$ and light $m_{S_3^{1/3}}$ to produce a sizable $W$ mass correction. 
In Fig. \ref{fig:mW:contour}, we show the parameter space allowed by the CDF $W$ mass measurement and the oblique parameters. For the pure $(X,T,B)_{L,R}$ case, we find that the $m_T$ should be at least 3.8TeV when $s_L=0.05$. For the pure $S_3$ case, we find that the $|\widetilde{\lambda}_{\phi S_3}|$ should be at least 2.3 when $m_{S_3^{1/3}}=1.5\mr{TeV}$. The $|\Delta m|$ lies around 25GeV, which is almost independent of $m_{S_3^{1/3}}$. \footnote{This is similar to the estimation in Ref. \cite{Athron:2022qpo}, in which the authors studied the singlet LQ $S_1$ and triplet LQ $S_3$ extended model. The $\Delta m$ in their work denotes the mass difference between $S_1$ and $S_3^{1/3}$.}

\begin{figure}[!h]
\begin{center}
\includegraphics[scale=0.4]{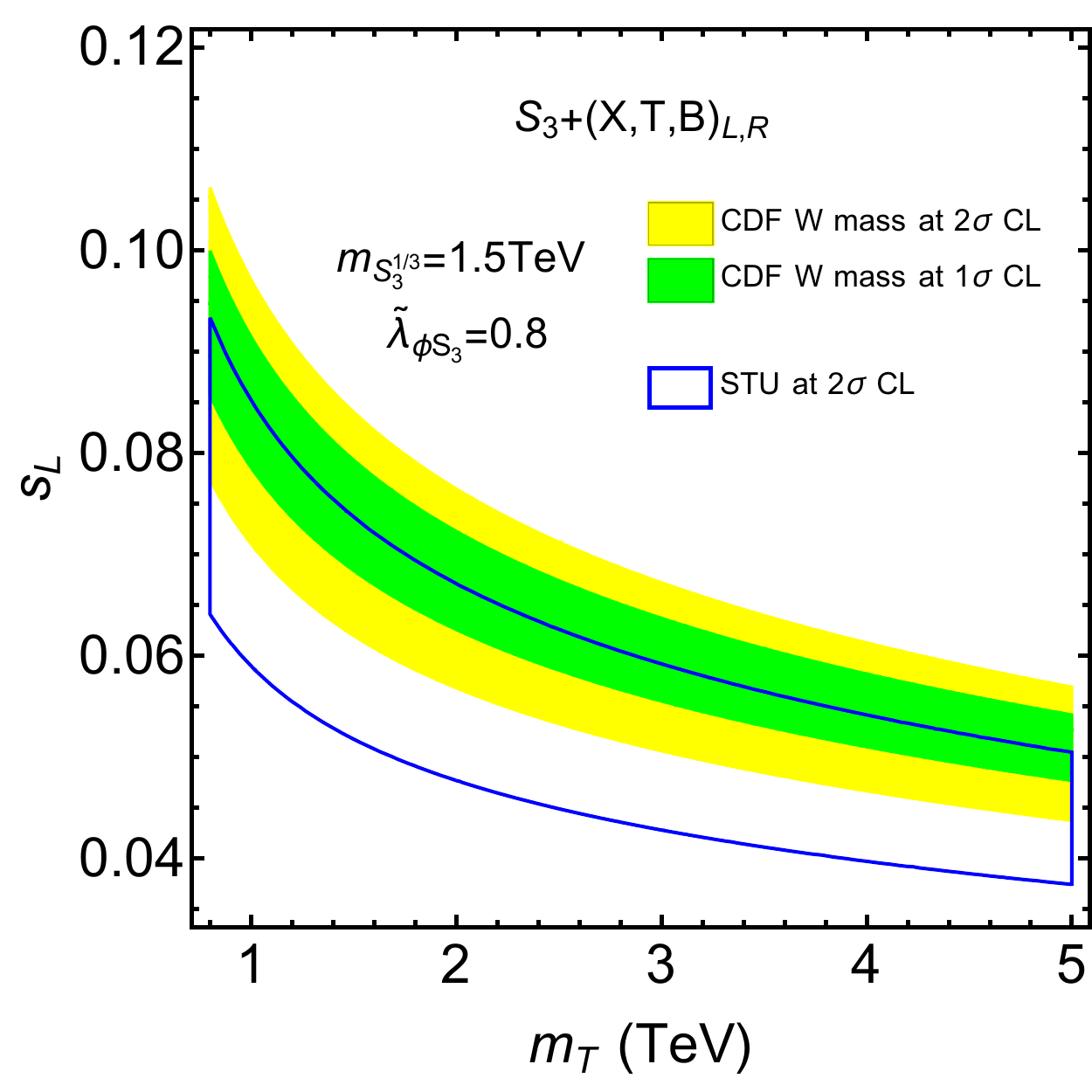}\qquad\qquad\qquad
\includegraphics[scale=0.4]{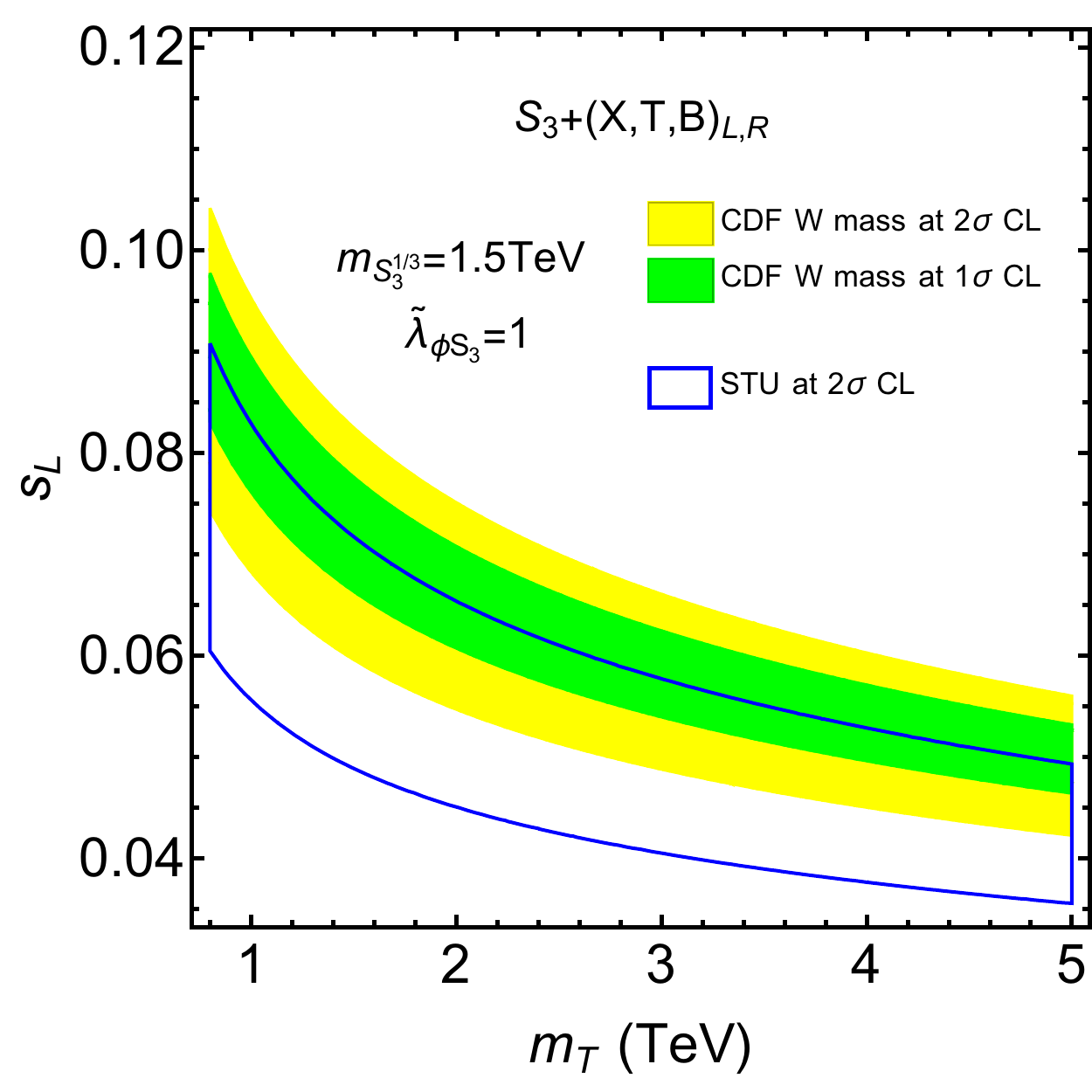}\\
\includegraphics[scale=0.4]{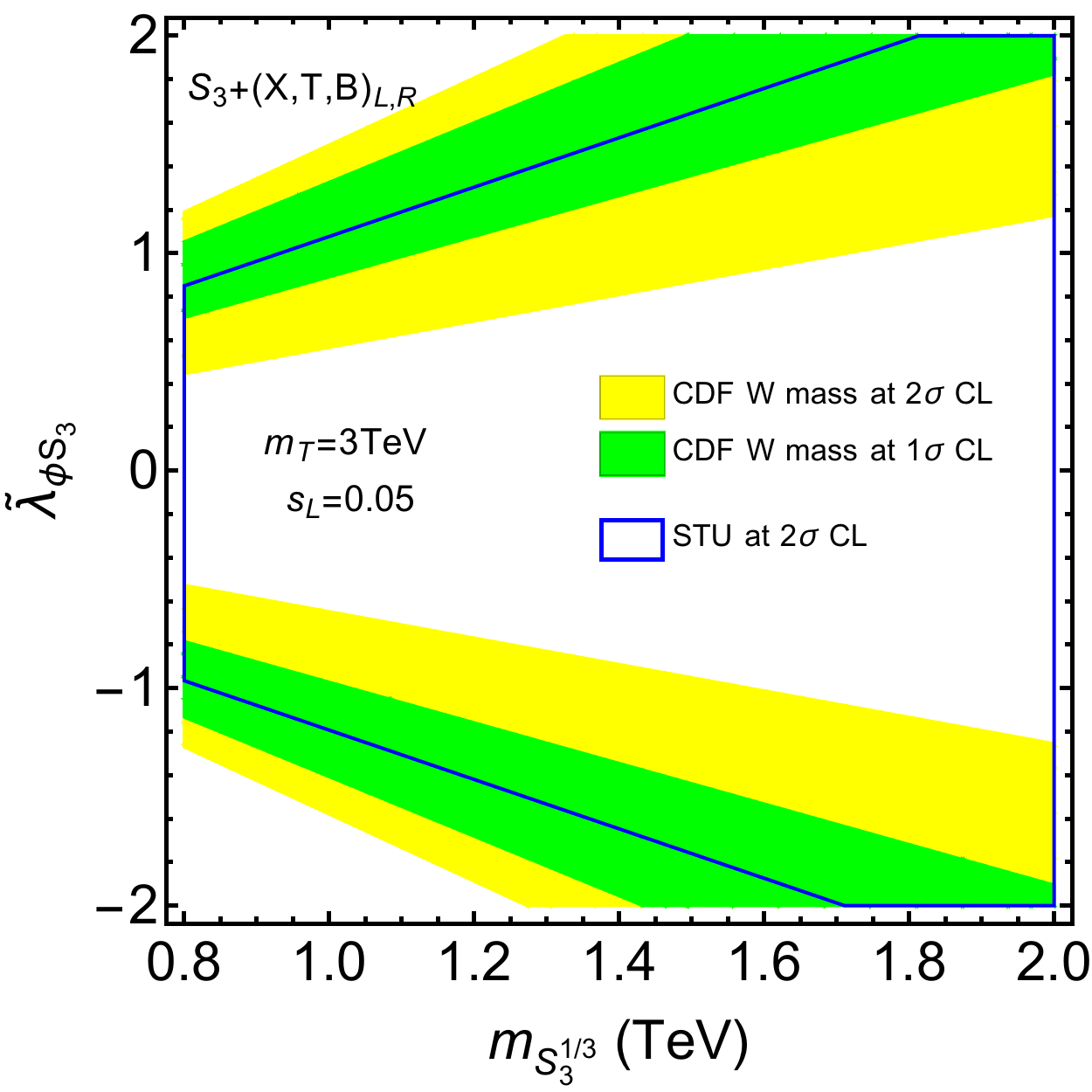}\qquad\qquad\qquad
\includegraphics[scale=0.4]{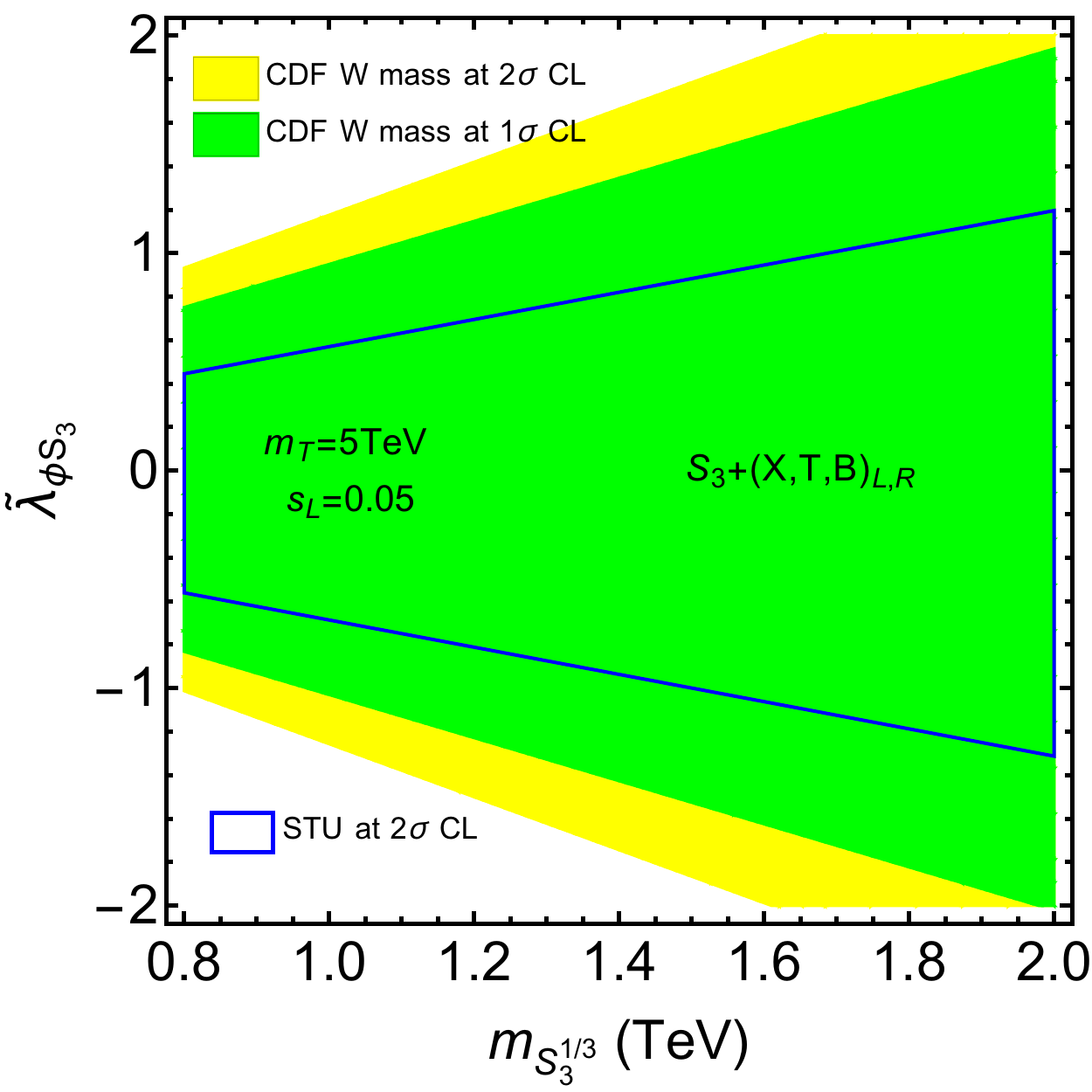}
\caption{The CDF $W$ mass allowed regions in the plane of $m_T-s_L$ for the scenarios $m_{S_3^{1/3}}=1.5\mr{TeV},\widetilde{\lambda}_{\phi S_3}=0.8$ (upper left) and $m_{S_3^{1/3}}=1.5\mr{TeV},\widetilde{\lambda}_{\phi S_3}=1$ (upper right). The CDF $W$ mass allowed regions in the plane of $m_{S_3^{1/3}}-\widetilde{\lambda}_{\phi S_3}$ for the scenarios $m_T=3\mr{TeV},s_L=0.05$ (lower left) and $m_T=5\mr{TeV},s_L=0.05$ (lower right). The blue line enclosed area is bounded by the $S,T,U$ parameters at $2\sigma$ CL.}\label{fig:mW:S3XTB}
\end{center}
\end{figure}

In Fig. \ref{fig:mW:S3XTB}, we consider the contributions from $(X,T,B)_{L,R}$ and $S_3$ at the same time. In the two plots above, we show the $W$ mass allowed regions in the plane of $m_T-s_L$ with fixed $m_{S_3^{1/3}}$ and $\widetilde{\lambda}_{\phi S_3}$. For the scenarios $m_{S_3^{1/3}}=1.5\mr{TeV}$ and $\widetilde{\lambda}_{\phi S_3}=0.8,1$, we find that the lower limit of $m_T$ can be decreased to 3.1TeV and 2.7TeV when $s_L=0.05$. In the two plots below, we show the $W$ mass allowed regions in the plane of $m_{S_3^{1/3}}-\widetilde{\lambda}_{\phi S_3}$ with fixed $m_T$ and $s_L$. For the scenario $s_L=0.05$ and $m_T=3\mr{TeV}$, we find that the lower limit of $|\widetilde{\lambda}_{\phi S_3}|$ can be decreased to 0.9 when $m_{S_3^{1/3}}=1.5\mr{TeV}$. For the scenario $s_L=0.05$ and $m_T=5\mr{TeV}$, the $\widetilde{\lambda}_{\phi S_3}$ can be zero because the pure $(X,T,B)_{L,R}$ is sufficient to produce the $W$ mass correction.

\begin{figure}[!htb]
\begin{center}
\includegraphics[scale=0.4]{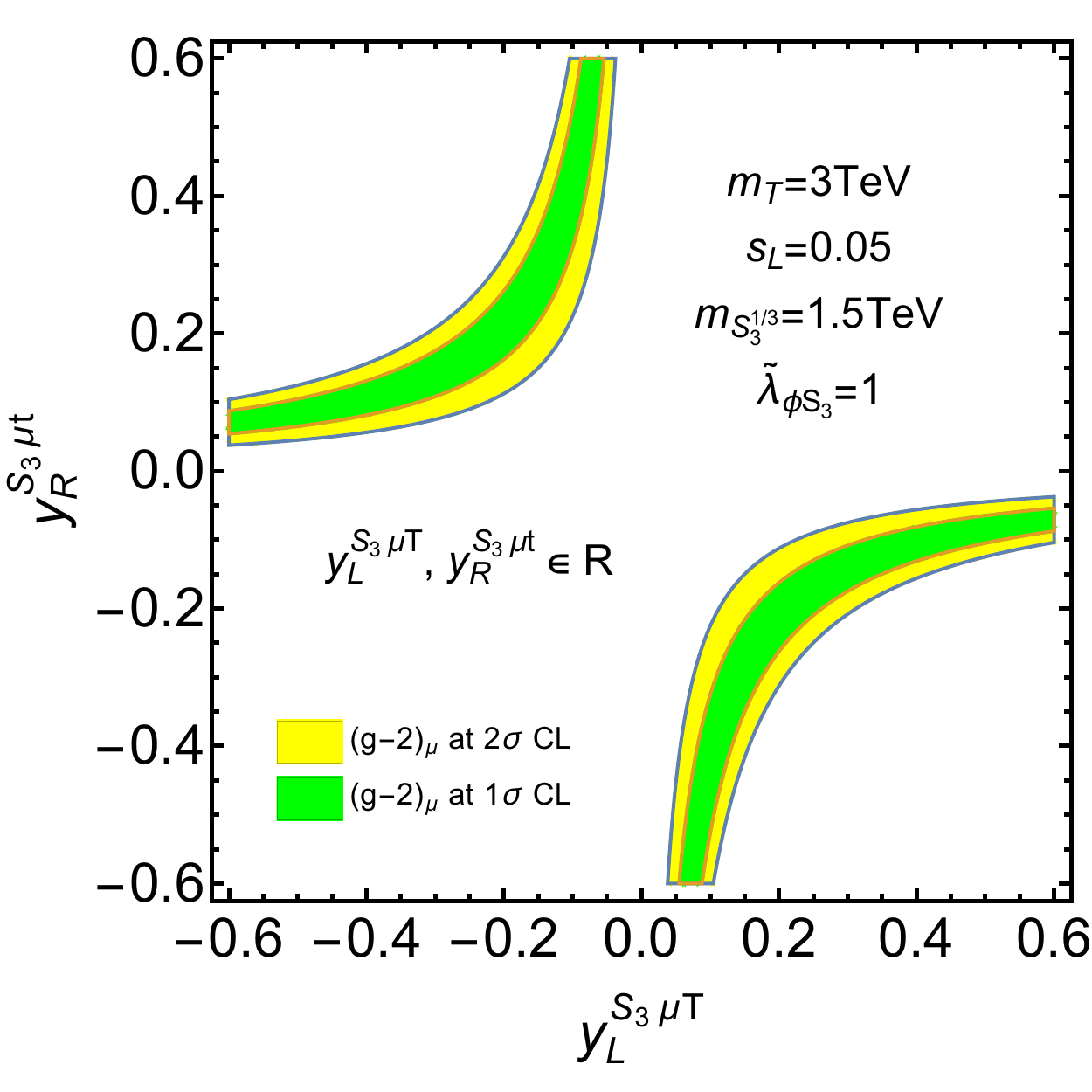}\qquad\qquad\qquad
\includegraphics[scale=0.4]{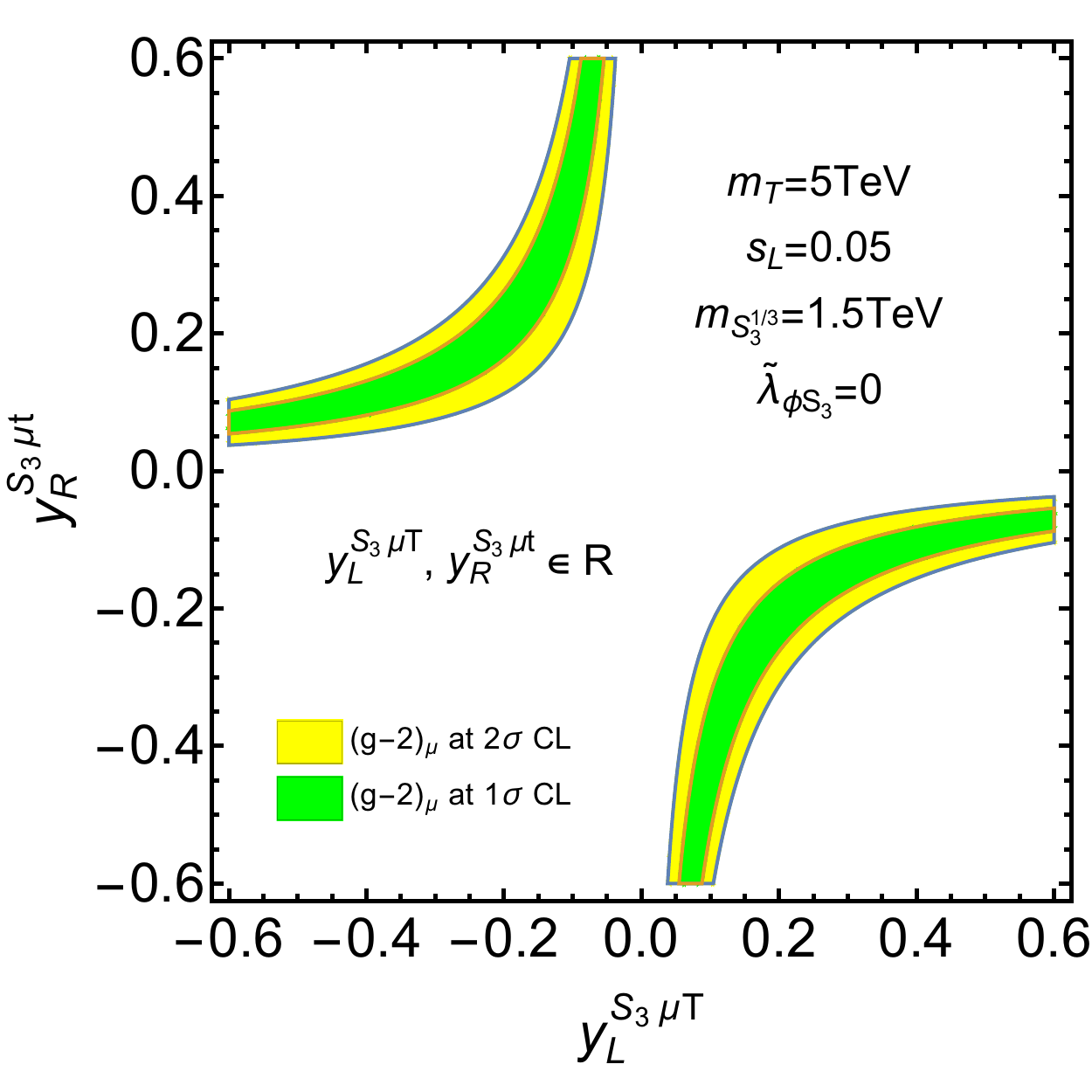}
\caption{The regions allowed by $(g-2)_{\mu}$ at $1\sigma$ (green) and $2\sigma$ (yellow) CLs, respectively. Here, we include the full contributions besides the chirally enhanced parts.}\label{fig:g-2:yLTyRt}
\end{center}
\end{figure}

Moreover, the $S_3+(X,T,B)_{L,R}$ model can also explain the $(g-2)_{\mu}$ anomaly. In our previous paper \cite{He:2021yck}, we took the LQs to be the same mass ($\widetilde{\lambda}_{\phi S_3}=0$). Here, we also consider the LQ mass differences, which only lead to small effects. Based on the previous $W$ mass numerical analysis, we choose two benchmark points $m_T=3\mr{TeV},s_L=0.05,m_{S_3^{1/3}}=1.5\mr{TeV},\widetilde{\lambda}_{\phi S_3}=1$ and $m_T=5\mr{TeV},s_L=0.05,m_{S_3^{1/3}}=1.5\mr{TeV},\widetilde{\lambda}_{\phi S_3}=0$. Under the first benchmark point, the leading order numerical result of $\Delta a_{\mu}$ is $-0.5914\times10^{-7}\mr{Re}[y_L^{S_3\mu T}(y_R^{S_3\mu t})^\ast]$, which constrains the $\mr{Re}[y_L^{S_3\mu T}(y_R^{S_3\mu t})^\ast]$ to be in the ranges $(-0.052,-0.032)$ and $(-0.062,-0.022)$ roughly at $1\sigma$ and $2\sigma$ CLs, respectively. Under the second benchmark point, the leading order numerical result of $\Delta a_{\mu}$ is $-0.4542\times10^{-7}\mr{Re}[y_L^{S_3\mu T}(y_R^{S_3\mu t})^\ast]$, which constrains the $\mr{Re}[y_L^{S_3\mu T}(y_R^{S_3\mu t})^\ast]$ to be in the ranges $(-0.068,-0.042)$ and $(-0.081,-0.029)$ roughly at $1\sigma$ and $2\sigma$ CLs, respectively. In Fig. \ref{fig:g-2:yLTyRt}, we show the regions allowed by $(g-2)_{\mu}$ in the plane of $y_L^{S_3\mu T}-y_R^{S_3\mu t}$.

\section{Summary and conclusions}\label{sec:summary}
We consider the $(X,T,B)_{L,R}$ and $S_3$ extended model to explain the $W$ boson mass anomaly. The mass splittings of VLQ are from the mixing with SM quark, and the mass splittings of LQ can be generated through the interaction with SM Higgs. For the VLQ oblique parameter corrections, some papers adopt the existing formulae directly without any examination, which are based on the singlet and doublet properties. In this paper, we obtain the complete VLQ and LQ contributions to the $S$, $T$, and $U$ parameters. As we know, the direct search experiments push the VLQ and LQ mass lower limits around TeV. We also consider the constraints from electroweak precision measurements, of which the $Zbb$ coupling imposes the strong bound $s_L\lesssim0.05$. For the pure $(X,T,B)_{L,R}$ model, the $m_T$ should be as heavy as 4TeV for $s_L=0.05$. For the pure $S_3$ model, it requires $|\widetilde{\lambda}_{\phi S_3}|\sim2$ for $m_{S_3^{1/3}}=1.5\mr{TeV}$. For the $S_3+(X,T,B)_{L,R}$ model, we find that the $W$ boson mass and $(g-2)_{\mu}$ anomalies can be explained simultaneously. Because the $W$ mass corrections can be shared by the VLQ and LQ, it allows lower $m_T$ and smaller $|\widetilde{\lambda}_{\phi S_3}|$. Depending on the choices of $m_T,s_L,m_{S_3^{1/3}}$, the $(g-2)_{\mu}$ anomaly can also be explained when $\mr{Re}[y_L^{S_3\mu T}(y_R^{S_3\mu t})^\ast]$ ranges from $\sim\mc{O}(-0.1)$ to $\mc{O}(-0.01)$.
\begin{acknowledgements}
We would like to thank the $W$ mass seminar held by the ``All Things EFT committee" and the seminar held in China (\href{https://indico.ihep.ac.cn/event/16155}{https://indico.ihep.ac.cn/event/16155}), from which we benefit a lot. We are grateful to Hiroshi Okada, Junjie Cao, Haiying Cai, and Rui Zhang for helpful discussions. This research was supported by an appointment to the Young Scientist Training Program at the APCTP through the Science and Technology Promotion Fund and Lottery Fund of the Korean Government. This was also supported by the Korean Local Governments-Gyeongsangbuk-do Province and Pohang City.
\end{acknowledgements}
\appendix

\section{Properties and expansions of the $(X,T,B)_{L,R}$ related parameters}\label{app:VLQ:para}
The relations between the VLQ parameters:
\begin{align}
&\tan\theta_R^t=\frac{m_t}{m_T}\tan\theta_L^t,\qquad\tan\theta_R^b=\frac{m_b}{m_B}\tan\theta_L^b,\qquad\sin2\theta_L^b=\frac{\sqrt{2}(m_T^2-m_t^2)}{m_B^2-m_b^2}\sin2\theta_L^t,\nonumber\\
&\qquad\qquad\qquad M_T^2=m_T^2(c_L^t)^2+m_t^2(s_L^t)^2=m_B^2(c_L^b)^2+m_b^2(s_L^b)^2,\nonumber\\
&M_T=m_Tc_L^tc_R^t+m_ts_L^ts_R^t=m_Bc_L^bc_R^b+m_bs_L^bs_R^b=\frac{m_ts_L^t}{s_R^t}=\frac{m_Tc_L^t}{c_R^t}=\frac{m_bs_L^b}{s_R^b}=\frac{m_Bc_L^b}{c_R^b},\nonumber\\
&\qquad\qquad\qquad\sqrt{2}(m_Tc_R^ts_L^t-m_tc_L^ts_R^t)=m_Bc_R^bs_L^b-m_bc_L^bs_R^b.
\end{align}
Because of the approximations $m_b\ll m_t\ll m_T$ and $s_L^t\ll1$, we have the following results:
\begin{align}
&\qquad\qquad\qquad\qquad\theta_R^t\approx \frac{m_t}{m_T}\theta_L^t,\qquad\qquad m_X=M_T\approx m_T[1-\frac{1}{2}(1-\frac{m_t^2}{m_T^2})(\theta_L^t)^2],\nonumber\\
&\theta_L^b\approx \frac{\sqrt{2}(m_T^2-m_t^2)}{m_T^2-m_b^2}\theta_L^t\approx\sqrt{2}(1-\frac{m_t^2}{m_T^2})\theta_L^t,\qquad\theta_R^b\approx \frac{\sqrt{2}m_b(m_T^2-m_t^2)}{m_T(m_T^2-m_b^2)}\theta_L^t\approx\frac{\sqrt{2}m_b}{m_T}(1-\frac{m_t^2}{m_T^2})\theta_L^t,\nonumber\\
&m_B\approx m_T[1+\frac{(m_T^2-m_t^2)(m_T^2-2m_t^2+m_b^2)}{2m_T^2(m_T^2-m_b^2)}(\theta_L^t)^2]\approx m_T[1+\frac{1}{2}(1-\frac{m_t^2}{m_T^2})(1-\frac{2m_t^2}{m_T^2})(\theta_L^t)^2].
\end{align}
\section{LQ contributions to the oblique parameters}\label{app:LQ:ST}
First of all, let us define the $B_0$ function as
\begin{align}
B_0(p^2,m_1^2,m_2^2)\equiv\Delta_{\epsilon}-\int_0^1dx\log\frac{xm_1^2+(1-x)m_2^2-x(1-x)p^2}{\mu^2}.
\end{align}
In the above, the $\Delta_{\epsilon}$ is defined as $1/\epsilon-\gamma_E+\log4\pi$. Here, we adopt the dimensional regularization, and $D=4-2\epsilon$ is the space-time dimension. The $\gamma_E$ is Euler's constant, and $\mu$ is the renormalization scale. 

According to the LQ gauge interactions derived in Sec. \ref{sec:LQ:gauge}, the self energies of neutral gauge bosons are calculated as
\begin{align}
&\Pi_{\gamma\gamma}(p^2)=\frac{e^2N_C}{16\pi^2}\sum\limits_{S_3^i}Q_{S_3^i}^2[\frac{1}{3}(4m_{S_3^i}^2-p^2)B_0(p^2,m_{S_3^i}^2,m_{S_3^i}^2)-\frac{4}{3}m_{S_3^i}^2B_0(0,m_{S_3^i}^2,m_{S_3^i}^2)-\frac{2}{9}p^2],\nonumber\\
&\Pi_{\gamma Z}(p^2)=\frac{egN_C}{16\pi^2c_W}\sum\limits_{S_3^i}Q_{S_3^i}(I_3^{S_3^i}-Q_{S_3^i}s_W^2)[\frac{1}{3}(4m_{S_3^i}^2-p^2)B_0(p^2,m_{S_3^i}^2,m_{S_3^i}^2)-\frac{4}{3}m_{S_3^i}^2B_0(0,m_{S_3^i}^2,m_{S_3^i}^2)-\frac{2}{9}p^2],\nonumber\\
&\Pi_{ZZ}(p^2)=\frac{g^2N_C}{16\pi^2c_W^2}\sum\limits_{S_3^i}(I_3^{S_3^i}-Q_{S_3^i}s_W^2)^2[\frac{1}{3}(4m_{S_3^i}^2-p^2)B_0(p^2,m_{S_3^i}^2,m_{S_3^i}^2)-\frac{4}{3}m_{S_3^i}^2B_0(0,m_{S_3^i}^2,m_{S_3^i}^2)-\frac{2}{9}p^2],\nonumber\\
\end{align}
where $S_3^i=S_3^{4/3},S_3^{1/3},S_3^{-2/3}$. The $Q_{S_3^i}$ and $I_3^{S_3^i}$ label their electric charge and third component of weak isospin, which means $Q_{S_3^{4/3}}=4/3,Q_{S_3^{1/3}}=1/3,Q_{S_3^{-2/3}}=-2/3$ and $I_3^{S_3^{4/3}}=1,I_3^{S_3^{1/3}}=0,I_3^{S_3^{-2/3}}=-1$.

Then, the self energy of $W$ boson is calculated as
\begin{align}
&\Pi_{WW}(p^2)=\frac{g^2N_C}{16\pi^2}\Big\{-\frac{p^2}{9}[3B_0(p^2,m_{S_3^{4/3}}^2,m_{S_3^{1/3}}^2)+3B_0(p^2,m_{S_3^{-2/3}}^2,m_{S_3^{1/3}}^2)+4]\nonumber\\
&+\frac{2}{3}[(m_{S_3^{4/3}}^2+m_{S_3^{1/3}}^2)B_0(p^2,m_{S_3^{4/3}}^2,m_{S_3^{1/3}}^2)+(m_{S_3^{-2/3}}^2+m_{S_3^{1/3}}^2)B_0(p^2,m_{S_3^{-2/3}}^2,m_{S_3^{1/3}}^2)\nonumber\\
	&-m_{S_3^{4/3}}^2B_0(0,m_{S_3^{4/3}}^2,m_{S_3^{4/3}}^2)-m_{S_3^{-2/3}}^2B_0(0,m_{S_3^{-2/3}}^2,m_{S_3^{-2/3}}^2)-2m_{S_3^{1/3}}^2B_0(0,m_{S_3^{1/3}}^2,m_{S_3^{1/3}}^2)]\nonumber\\
&-\frac{1}{3p^2}[(m_{S_3^{4/3}}^2-m_{S_3^{1/3}}^2)^2B_0(p^2,m_{S_3^{4/3}}^2,m_{S_3^{1/3}}^2)+(m_{S_3^{-2/3}}^2-m_{S_3^{1/3}}^2)^2B_0(p^2,m_{S_3^{-2/3}}^2,m_{S_3^{1/3}}^2)\nonumber\\
	&-m_{S_3^{4/3}}^2(m_{S_3^{4/3}}^2-m_{S_3^{1/3}}^2)B_0(0,m_{S_3^{4/3}}^2,m_{S_3^{4/3}}^2)-m_{S_3^{-2/3}}^2(m_{S_3^{-2/3}}^2-m_{S_3^{1/3}}^2)B_0(0,m_{S_3^{-2/3}}^2,m_{S_3^{-2/3}}^2)\nonumber\\
	&+m_{S_3^{1/3}}^2(m_{S_3^{4/3}}^2+m_{S_3^{-2/3}}^2-2m_{S_3^{1/3}}^2)B_0(0,m_{S_3^{1/3}}^2,m_{S_3^{1/3}}^2)-(m_{S_3^{4/3}}^2-m_{S_3^{1/3}}^2)^2-(m_{S_3^{-2/3}}^2-m_{S_3^{1/3}}^2)^2]\Big\}.
\end{align}

Based on the exact expressions of $\Pi_{VV}(p^2)$, we can derive the $\Pi_{VV}(0)$ and $\Pi_{VV}'(0)\equiv d\Pi_{VV}(p^2)/dp^2|_{p^2=0}$ as follows:
\begin{align}\label{eqn:LQ:selfEapp}
&\Pi_{\gamma\gamma}(0)=\Pi_{\gamma Z}(0)=\Pi_{ZZ}(0)=0,\nonumber\\
&\Pi_{\gamma\gamma}'(0)=-\frac{e^2N_C}{48\pi^2}\sum\limits_{S_3^i}Q_{S_3^i}^2(\Delta_{\epsilon}-\log\frac{m_{S_3^i}^2}{\mu^2}),\nonumber\\
&\Pi_{\gamma Z}'(0)=-\frac{egN_C}{48\pi^2c_W}\sum\limits_{S_3^i}Q_{S_3^i}(I_3^{S_3^i}-Q_{S_3^i}s_W^2)(\Delta_{\epsilon}-\log\frac{m_{S_3^i}^2}{\mu^2}),\nonumber\\
&\Pi_{ZZ}'(0)=-\frac{g^2N_C}{48\pi^2c_W^2}\sum\limits_{S_3^i}(I_3^{S_3^i}-Q_{S_3^i}s_W^2)^2(\Delta_{\epsilon}-\log\frac{m_{S_3^i}^2}{\mu^2}),\nonumber\\
&\Pi_{WW}(0)=\frac{g^2N_C}{16\pi^2}[\frac{1}{2}(m_{S_3^{4/3}}^2+m_{S_3^{-2/3}}^2-2m_{S_3^{1/3}}^2)-\frac{m_{S_3^{1/3}}^2m_{S_3^{4/3}}^2}{m_{S_3^{1/3}}^2-m_{S_3^{4/3}}^2}\log\frac{m_{S_3^{1/3}}^2}{m_{S_3^{4/3}}^2}-\frac{m_{S_3^{1/3}}^2m_{S_3^{-2/3}}^2}{m_{S_3^{1/3}}^2-m_{S_3^{-2/3}}^2}\log\frac{m_{S_3^{1/3}}^2}{m_{S_3^{-2/3}}^2}],\nonumber\\
&\Pi_{WW}'(0)=\frac{g^2N_C}{16\pi^2}[-\frac{2}{3}(\Delta_{\epsilon}-\log\frac{m_{S_3^{1/3}}^2}{\mu^2})-\frac{4}{9}-\frac{m_{S_3^{4/3}}^4+m_{S_3^{1/3}}^4-14m_{S_3^{4/3}}^2m_{S_3^{1/3}}^2}{18(m_{S_3^{4/3}}^2-m_{S_3^{1/3}}^2)^2}-\frac{m_{S_3^{-2/3}}^4+m_{S_3^{1/3}}^4-14m_{S_3^{-2/3}}^2m_{S_3^{1/3}}^2}{18(m_{S_3^{-2/3}}^2-m_{S_3^{1/3}}^2)^2}\nonumber\\
	&+\frac{m_{S_3^{4/3}}^4(m_{S_3^{4/3}}^2-3m_{S_3^{1/3}}^2)}{3(m_{S_3^{4/3}}^2-m_{S_3^{1/3}}^2)^3}\log\frac{m_{S_3^{4/3}}^2}{m_{S_3^{1/3}}^2}+\frac{m_{S_3^{-2/3}}^4(m_{S_3^{-2/3}}^2-3m_{S_3^{1/3}}^2)}{3(m_{S_3^{-2/3}}^2-m_{S_3^{1/3}}^2)^3}\log\frac{m_{S_3^{-2/3}}^2}{m_{S_3^{1/3}}^2}].
\end{align}
The $S$, $T$, and $U$ parameters are defined as \cite{Peskin:1990zt, Peskin:1991sw, Maksymyk:1993zm}
\begin{align}\label{eqn:general:ST}
&\frac{\alpha S}{4s_W^2c_W^2}\equiv\frac{\Pi_{ZZ}(m_Z^2)-\Pi_{ZZ}(0)}{m_Z^2}-\frac{c_W^2-s_W^2}{s_Wc_W}\Pi_{\gamma Z}'(0)-\Pi_{\gamma\gamma}'(0)=\Pi_{ZZ}'(0)-\frac{c_W^2-s_W^2}{s_Wc_W}\Pi_{\gamma Z}'(0)-\Pi_{\gamma\gamma}'(0),\nonumber\\
&\qquad\qquad\qquad\qquad\qquad\qquad\qquad\alpha T\equiv\frac{\Pi_{WW}(0)}{m_W^2}-\frac{\Pi_{ZZ}(0)}{m_Z^2},\nonumber\\
&\frac{\alpha U}{4s_W^2}\equiv\frac{\Pi_{WW}(m_W^2)-\Pi_{WW}(0)}{m_W^2}-c_W^2\frac{\Pi_{ZZ}(m_Z^2)-\Pi_{ZZ}(0)}{m_Z^2}-2s_Wc_W\Pi_{\gamma Z}'(0)-s_W^2\Pi_{\gamma\gamma}'(0)\nonumber\\
	&\qquad=\Pi_{WW}'(0)-c_W^2\Pi_{ZZ}'(0)-2s_Wc_W\Pi_{\gamma Z}'(0)-s_W^2\Pi_{\gamma\gamma}'(0).
\end{align}
When we adopt Eq. \eqref{eqn:LQ:selfEapp} in the above definitions, the LQ contributions to the $S$, $T$, and $U$ parameters are calculated in the following explicit forms:
\begin{align}
&\Delta S^{S_3}=-\frac{N_c}{9\pi}\log\frac{m_{S_3^{4/3}}^2}{m_{S_3^{-2/3}}^2},\qquad \Delta T^{S_3}=\frac{N_c}{8\pi m_W^2s_W^2}[\theta_+(m_{S_3^{4/3}}^2,m_{S_3^{1/3}}^2)+\theta_+(m_{S_3^{-2/3}}^2,m_{S_3^{1/3}}^2)],\nonumber\\
&\Delta U^{S_3}=\frac{N_C}{\pi}[-\frac{4}{9}-\frac{m_{S_3^{4/3}}^4+m_{S_3^{1/3}}^4-14m_{S_3^{4/3}}^2m_{S_3^{1/3}}^2}{18(m_{S_3^{4/3}}^2-m_{S_3^{1/3}}^2)^2}-\frac{m_{S_3^{-2/3}}^4+m_{S_3^{1/3}}^4-14m_{S_3^{-2/3}}^2m_{S_3^{1/3}}^2}{18(m_{S_3^{-2/3}}^2-m_{S_3^{1/3}}^2)^2}\nonumber\\
	&+\frac{m_{S_3^{1/3}}^4(-3m_{S_3^{4/3}}^2+m_{S_3^{1/3}}^2)}{3(m_{S_3^{4/3}}^2-m_{S_3^{1/3}}^2)^3}\log\frac{m_{S_3^{4/3}}^2}{m_{S_3^{1/3}}^2}+\frac{m_{S_3^{1/3}}^4(-3m_{S_3^{-2/3}}^2+m_{S_3^{1/3}}^2)}{3(m_{S_3^{-2/3}}^2-m_{S_3^{1/3}}^2)^3}\log\frac{m_{S_3^{-2/3}}^2}{m_{S_3^{1/3}}^2}].
\end{align}
As we can see, the divergence and scale $\mu$ are exactly cancelled in the oblique parameters.
\section{VLQ contributions to the oblique parameters}\label{app:VLQ:ST}
Because the triplet VLQ is involved, we can not adopt the formulae of $S$ and $T$ parameters in Ref. \cite{Lavoura:1992np} simply, in which some calculations are based on the singlet and doublet properties. In this section, we present a detailed deduction.

In general, let us denote the quark interactions with gauge bosons $V_1$ and $V_2$ as $\bar{f_i}\gamma_{\mu}(g_{V_1}^{ij}+g_{A_1}^{ij}\gamma^5)f_jV_1^{\mu}+\bar{f_i}\gamma_{\mu}(g_{V_2}^{ij}+g_{A_2}^{ij}\gamma^5)f_jV_2^{\mu}$. Here, the masses of $f_i$ and $f_j$ are labelled as $m_i$ and $m_j$. Thus, the self energy of $V_1-V_2$ is calculated as \cite{Lavoura:1992np}
\begin{align}\label{eqn:VLQ:self}
&\Pi_{V_1V_2}(0)=\frac{N_C}{16\pi^2}\big\{(g_{V_1}^{ij}g_{V_2}^{ij}+g_{A_1}^{ij}g_{A_2}^{ij})[2(m_i^2+m_j^2)\Delta_{\epsilon}-2(m_i^2\log\frac{m_i^2}{\mu^2}+m_j^2\log\frac{m_j^2}{\mu^2})+\theta_+(m_i^2,m_j^2)]\nonumber\\
&+(g_{V_1}^{ij}g_{V_2}^{ij}-g_{A_1}^{ij}g_{A_2}^{ij})[-4m_im_j\Delta_{\epsilon}+2m_im_j\log\frac{m_i^2m_j^2}{\mu^4}+\theta_-(m_i^2,m_j^2)]\big\}.
\end{align}
Besides, the first derivative of $V_1-V_2$ self energy is calculated as \cite{Lavoura:1992np}
\begin{align}\label{eqn:VLQ:dself}
&\Pi_{V_1V_2}'(0)\equiv\frac{d\Pi_{V_1V_2}(p^2)}{dp^2}|_{p^2=0}=\frac{N_C}{4\pi^2}\big\{(g_{V_1}^{ij}g_{V_2}^{ij}+g_{A_1}^{ij}g_{A_2}^{ij})[-\frac{1}{3}\Delta_{\epsilon}+\frac{1}{6}+\frac{1}{6}\log\frac{m_i^2m_j^2}{\mu^4}-\frac{1}{2}\chi_+(m_i^2,m_j^2)]\nonumber\\
&+(g_{V_1}^{ij}g_{V_2}^{ij}-g_{A_1}^{ij}g_{A_2}^{ij})[-\frac{m_i^2+m_j^2}{12m_im_j}-\frac{1}{2}\chi_-(m_i^2,m_j^2)]\big\}.
\end{align}
In Ref. \cite{Lavoura:1992np}, the derivation of $S$ and $T$ parameters depends on the following relations:
\begin{align}
&(U^\alpha)^2=U^\alpha,\quad(D^\alpha)^2=D^\alpha,\quad D^LM_dD^R=(V^L)^{\dag}M_uV^R,\quad U^LM_uU^R=V^LM_d(V^R)^{\dag}.
\end{align}
While, they are only valid for the singlet and doublet VLQs. For the $(X,T,B)$ case, they no longer hold. In fact, the cancellation of divergence is guaranteed by the relations in Eq. \eqref{eqn:HYukawa:relations}.

For compactness and simplicity, let us reformulate the VLQ gauge interactions in Sec. \ref{sec:VLQ:gauge} with the matrix form. The gauge interactions with $W$ boson can be written as
\begin{align}
&\frac{g}{\sqrt{2}}W_{\mu}^+\Big\{\overline{X_L}\gamma^{\mu}V_L^{Xt}\left[\begin{array}{c}t_L\\T_L\end{array}\right]+\overline{X_R}\gamma^{\mu}V_R^{Xt}\left[\begin{array}{c}t_R\\T_R\end{array}\right]\nonumber\\
&+(\overline{t_L},\overline{T_L})\gamma^{\mu}V_L^{tb}\left[\begin{array}{c}b_L\\B_L\end{array}\right]+(\overline{t_R},\overline{T_R})\gamma^{\mu}V_R^{tb}\left[\begin{array}{c}b_R\\B_R\end{array}\right]\Big\}+\mathrm{h.c.}.
\end{align}
Similarly, the gauge interactions with $Z$ boson can be written as
\begin{align}
&\frac{g}{2c_W}Z_{\mu}\Big\{\overline{X_L}\gamma^{\mu}(U_L^X-2Q_Xs_W^2)X_L+\overline{X_R}\gamma^{\mu}(U_R^X-2Q_Xs_W^2)X_R+(\overline{t_L},\overline{T_L})\gamma^{\mu}(U_L^t-2Q_ts_W^2)\left[\begin{array}{c}t_L\\T_L\end{array}\right]\nonumber\\
&+(\overline{t_R},\overline{T_R})\gamma^{\mu}(U_R^t-2Q_ts_W^2)\left[\begin{array}{c}t_R\\T_R\end{array}\right]-(\overline{b_L},\overline{B_L})\gamma^{\mu}(U_L^b+2Q_bs_W^2)\left[\begin{array}{c}b_L\\B_L\end{array}\right]-(\overline{b_R},\overline{B_R})\gamma^{\mu}(U_R^b+2Q_bs_W^2)\left[\begin{array}{c}b_R\\B_R\end{array}\right]\Big\}.
\end{align}
As to the gauge interactions with photon, it has the trivial from $eQ_f\bar{f}\gamma^{\mu}fA_{\mu}$. In the above, the $V$ and $U$ matrices are given as
\begin{align}
&\qquad\qquad\qquad  V_L^{Xt}=\sqrt{2}(s_L^t,-c_L^t),\qquad V_R^{Xt}=\sqrt{2}(s_R^t,-c_R^t),\nonumber\\
&V_L^{tb}=\left[\begin{array}{cc}c_L^tc_L^b+\sqrt{2}s_L^ts_L^b&c_L^ts_L^b-\sqrt{2}s_L^tc_L^b\\s_L^tc_L^b-\sqrt{2}c_L^ts_L^b&s_L^ts_L^b+\sqrt{2}c_L^tc_L^b\end{array}\right],\qquad V_R^{tb}=\left[\begin{array}{cc}\sqrt{2}s_R^ts_R^b&-\sqrt{2}s_R^tc_R^b\\-\sqrt{2}c_R^ts_R^b&\sqrt{2}c_R^tc_R^b\end{array}\right],
\end{align}
and
\begin{align}
&U_L^X=U_R^X=2,\qquad U_L^t=\left[\begin{array}{cc}(c_L^t)^2&s_L^tc_L^t\\s_L^tc_L^t&(s_L^t)^2\end{array}\right],\qquad U_R^t=\left[\begin{array}{cc}0&0\\0&0\end{array}\right],\nonumber\\
&U_L^b=\left[\begin{array}{cc}1+(s_L^b)^2&-s_L^bc_L^b\\-s_L^bc_L^b&1+(c_L^b)^2\end{array}\right],\qquad U_R^b=\left[\begin{array}{cc}2(s_R^b)^2&-2s_R^bc_R^b\\-2s_R^bc_R^b&2(c_R^b)^2\end{array}\right].
\end{align}
The $U$ and $V$ matrices can be correlated through the following identities: 
\begin{align}
U_{L/R}^X=V_{L/R}^{Xt}(V_{L/R}^{Xt})^{\dag},\qquad U_{L/R}^t=V_{L/R}^{tb}(V_{L/R}^{tb})^{\dag}-(V_{L/R}^{Xt})^{\dag}V_{L/R}^{Xt},\qquad U_{L/R}^b=(V_{L/R}^{tb})^{\dag}V_{L/R}^{tb}.
\end{align}
\subsection{Derivation of the $T$ parameter}
According to the Eq. \eqref{eqn:VLQ:self}, the self energy consists of the $\theta_{\pm}$ and non-$\theta_{\pm}$ parts. Now, let us consider the non-$\theta_{\pm}$ part. Based on the definition in Eq. \eqref{eqn:general:ST}, it can be calculated as
\begin{align}
&\alpha T_{non-\theta_{\pm}}^{XTB}=\frac{N_Cg^2}{32\pi^2m_W^2}\Big\{[V_L^{Xt}(V_L^{Xt})^{\dag}-U_L^XU_L^X+V_R^{Xt}(V_R^{Xt})^{\dag}-U_R^XU_R^X]m_X^2(\Delta_{\epsilon}-\log\frac{m_X^2}{\mu^2})\nonumber\\
&+[V_L^{Xt}.M_u.(V_R^{Xt})^{\dag}-U_L^Xm_XU_R^X+V_R^{Xt}.M_u.(V_L^{Xt})^{\dag}-U_R^Xm_XU_L^X]m_X(-\Delta_{\epsilon}+\log\frac{m_X^2}{\mu^2})\nonumber\\
&+\mr{Tr}[\big(V_L^{tb}(V_L^{tb})^{\dag}+(V_L^{Xt})^{\dag}V_L^{Xt}-U_L^tU_L^t+V_R^{tb}(V_R^{tb})^{\dag}+(V_R^{Xt})^{\dag}V_R^{Xt}-U_R^tU_R^t\big).M_u^2.(\Delta_{\epsilon}-\log\frac{M_u^2}{\mu^2})]\nonumber\\
&+\mr{Tr}[\big(V_L^{tb}.M_d.(V_R^{tb})^{\dag}+(V_L^{Xt})^{\dag}m_XV_R^{Xt}-U_L^t.M_u.U_R^t+V_R^{tb}.M_d.(V_L^{tb})^{\dag}+(V_R^{Xt})^{\dag}m_XV_L^{Xt}-U_R^t.M_u.U_L^t\big).\nonumber\\
&M_u.(-\Delta_{\epsilon}+\log\frac{M_u^2}{\mu^2})]+\mr{Tr}[\big((V_L^{tb})^{\dag}V_L^{tb}-U_L^bU_L^b+(V_R^{tb})^{\dag}V_R^{tb}-U_R^bU_R^b\big).M_d^2.(\Delta_{\epsilon}-\log\frac{M_d^2}{\mu^2})]\nonumber\\
&+\mr{Tr}[\big((V_L^{tb})^{\dag}.M_u.V_R^{tb}-U_L^b.M_d.U_R^b+(V_R^{tb})^{\dag}.M_u.V_L^{tb}-U_R^b.M_d.U_L^b\big).M_d.(-\Delta_{\epsilon}+\log\frac{M_d^2}{\mu^2})]\Big\}\nonumber\\
&=\frac{N_Cg^2}{16\pi^2m_W^2}\Big\{m_t(\Delta_{\epsilon}-\log\frac{m_t^2}{\mu^2})[2m_t\big((s_L^t)^2+(s_R^t)^2\big)+\sqrt{2}c_L^ts_R^t(m_Bc_R^bs_L^b-m_bc_L^bs_R^b)-4m_Xs_L^ts_R^t]\nonumber\\
&+m_T(\Delta_{\epsilon}-\log\frac{m_T^2}{\mu^2})[2m_T\big((c_L^t)^2+(c_R^t)^2\big)-\sqrt{2}s_L^tc_R^t(m_Bc_R^bs_L^b-m_bc_L^bs_R^b)-4m_Xc_L^tc_R^t]\nonumber\\
&+m_b(\Delta_{\epsilon}-\log\frac{m_b^2}{\mu^2})[m_b\big(-(s_L^b)^2+(s_R^b)^2\big)+\sqrt{2}c_L^bs_R^b(m_Tc_R^ts_L^t-m_tc_L^ts_R^t)]\nonumber\\
&+m_B(\Delta_{\epsilon}-\log\frac{m_B^2}{\mu^2})[m_B\big(-(c_L^b)^2+(c_R^b)^2\big)-\sqrt{2}c_R^bs_L^b(m_Tc_R^ts_L^t-m_tc_L^ts_R^t)]\Big\}=0,
\end{align}
where the two diagonal matrices $M_u$ and $M_d$ are defined as $M_u=\mr{Diag}\{m_t,m_T\}$ and $M_d=\mr{Diag}\{m_b,m_B\}$. We find that they are exactly cancelled, thus only the $\theta_{\pm}$ part can contribute to the $T$ parameter. Then, the $T$ parameter formula in Ref. \cite{Lavoura:1992np} still stands for the $(X,T,B)_{L,R}$ triplet case. 

For the $\Delta T^{XTB}$ parameter, it is computed as
\begin{align}\label{eqn:XTB:T}
&\Delta T^{XTB}=\frac{N_C}{16\pi s_W^2m_W^2}\Big\{2[(s_L^t)^2+(s_R^t)^2]\theta_+(m_X^2,m_t^2)+4s_L^ts_R^t\theta_-(m_X^2,m_t^2)\nonumber\\
&+2[(c_L^t)^2+(c_R^t)^2]\theta_+(m_X^2,m_T^2)+4c_L^tc_R^t\theta_-(m_X^2,m_T^2)\nonumber\\
&+[(c_L^tc_L^b+\sqrt{2}s_L^ts_L^b)^2+2(s_R^ts_R^b)^2-1]\theta_+(m_t^2,m_b^2)+2\sqrt{2}s_R^ts_R^b(c_L^tc_L^b+\sqrt{2}s_L^ts_L^b)\theta_-(m_t^2,m_b^2)\nonumber\\
&+[(c_L^ts_L^b-\sqrt{2}s_L^tc_L^b)^2+2(s_R^tc_R^b)^2]\theta_+(m_t^2,m_B^2)-2\sqrt{2}s_R^tc_R^b(c_L^ts_L^b-\sqrt{2}s_L^tc_L^b)\theta_-(m_t^2,m_B^2)\nonumber\\
&+[(s_L^tc_L^b-\sqrt{2}c_L^ts_L^b)^2+2(c_R^ts_R^b)^2]\theta_+(m_T^2,m_b^2)-2\sqrt{2}c_R^ts_R^b(s_L^tc_L^b-\sqrt{2}c_L^ts_L^b)\theta_-(m_T^2,m_b^2)\nonumber\\
&+[(s_L^ts_L^b+\sqrt{2}c_L^tc_L^b)^2+2(c_R^tc_R^b)^2]\theta_+(m_T^2,m_B^2)+2\sqrt{2}c_R^tc_R^b(s_L^ts_L^b+\sqrt{2}c_L^tc_L^b)\theta_-(m_T^2,m_B^2)\nonumber\\
&-(s_L^tc_L^t)^2\chi_+(m_t^2,m_T^2)-[(s_L^bc_L^b)^2+4(s_R^bc_R^b)^2]\theta_+(m_b^2,m_B^2)-4(s_L^bc_L^b)(s_R^bc_R^b)\theta_-(m_b^2,m_B^2)\Big\}.
\end{align}
Here, the $\theta_+$ function has been defined before, and the $\theta_-$ function is defined as
\begin{align}
\theta_-(y_1,y_2)\equiv2\sqrt{y_1y_2}[\frac{y_1+y_2}{y_1-y_2}\log\frac{y_1}{y_2}-2].
\end{align}
It is consistent with the result of $T$ parameter in Ref. \cite{Cai:2012ji}. 

\subsection{Derivation of the $S$ parameter}
If we adopt the $S$ parameter formula in Ref. \cite{Lavoura:1992np} naively, it will give the following result:
\begin{align}\label{eqn:XTB:SNaive}
&\Delta S_{wrong}^{XTB}=\frac{N_C}{2\pi}\Big\{2[(s_L^t)^2+(s_R^t)^2]\psi_+(m_X^2,m_t^2)+4s_L^ts_R^t\psi_-(m_X^2,m_t^2)\nonumber\\
&+2[(c_L^t)^2+(c_R^t)^2]\psi_+(m_X^2,m_T^2)+4c_L^tc_R^t\psi_-(m_X^2,m_T^2)\nonumber\\
&+[(c_L^tc_L^b+\sqrt{2}s_L^ts_L^b)^2+2(s_R^ts_R^b)^2-1]\psi_+(m_t^2,m_b^2)+2\sqrt{2}s_R^ts_R^b(c_L^tc_L^b+\sqrt{2}s_L^ts_L^b)\psi_-(m_t^2,m_b^2)\nonumber\\
&+[(c_L^ts_L^b-\sqrt{2}s_L^tc_L^b)^2+2(s_R^tc_R^b)^2]\psi_+(m_t^2,m_B^2)-2\sqrt{2}s_R^tc_R^b(c_L^ts_L^b-\sqrt{2}s_L^tc_L^b)\psi_-(m_t^2,m_B^2)\nonumber\\
&+[(s_L^tc_L^b-\sqrt{2}c_L^ts_L^b)^2+2(c_R^ts_R^b)^2]\psi_+(m_T^2,m_b^2)-2\sqrt{2}c_R^ts_R^b(s_L^tc_L^b-\sqrt{2}c_L^ts_L^b)\psi_-(m_T^2,m_b^2)\nonumber\\
&+[(s_L^ts_L^b+\sqrt{2}c_L^tc_L^b)^2+2(c_R^tc_R^b)^2]\psi_+(m_T^2,m_B^2)+2\sqrt{2}c_R^tc_R^b(s_L^ts_L^b+\sqrt{2}c_L^tc_L^b)\psi_-(m_T^2,m_B^2)\nonumber\\
&-(s_L^tc_L^t)^2\chi_+(m_t^2,m_T^2)-[(s_L^bc_L^b)^2+4(s_R^bc_R^b)^2]\chi_+(m_b^2,m_B^2)-4(s_L^bc_L^b)(s_R^bc_R^b)\chi_-(m_b^2,m_B^2)\Big\}.
\end{align}
In the above, the functions $\psi_{\pm}$ and $\chi_{\pm}$ are defined as
\begin{align}
&\psi_+(y_1,y_2)\equiv\frac{1}{3}-\frac{1}{9}\log\frac{y_1}{y_2},\qquad\qquad\qquad\psi_-(y_1,y_2)\equiv-\frac{y_1+y_2}{6\sqrt{y_1y_2}},\nonumber\\
&\chi_+(y_1,y_2)\equiv\frac{5(y_1^2+y_2^2)-22y_1y_2}{9(y_1-y_2)^2}+\frac{3y_1y_2(y_1+y_2)-y_1^3-y_2^3}{3(y_1-y_2)^3}\log\frac{y_1}{y_2},\nonumber\\
&\chi_-(y_1,y_2)\equiv-\sqrt{y_1y_2}[\frac{y_1+y_2}{6y_1y_2}-\frac{y_1+y_2}{(y_1-y_2)^2}+\frac{2y_1y_2}{(y_1-y_2)^3}\log\frac{y_1}{y_2}].
\end{align}
While, it is not solid because the $S$ parameter formula in Ref. \cite{Lavoura:1992np} relies on the singlet and doublet representations, which should be reconsidered for the $(X,T,B)_{L,R}$ triplet \footnote{I would like to thank Haiying Cai for talking about this.}.

According to the Eq. \eqref{eqn:VLQ:dself}, the first derivative of self energy consists of the $\chi_{\pm}$ and non-$\chi_{\pm}$ parts. Now, let us consider the non-$\chi_{\pm}$ part. Based on the definition in Eq. \eqref{eqn:general:ST}, it can be calculated as
\begin{align}
&\quad\frac{\alpha S_{non-\chi_{\pm}}^{XTB}}{4s_W^2c_W^2}\nonumber\\
&=\frac{N_Cg^2}{96\pi^2c_W^2}\Big\{[U_L^XU_L^X+U_R^XU_R^X-2Q_X(U_L^X+U_R^X)](-\Delta_{\epsilon}+\log\frac{m_X^2}{\mu^2})+\frac{U_L^XU_L^X+U_R^XU_R^X}{2}-U_L^XU_R^X\nonumber\\
&+\mr{Tr}[\big(U_L^tU_L^t+U_R^tU_R^t-2Q_t(U_L^t+U_R^t)\big).(-\Delta_{\epsilon}+\log\frac{M_u^2}{\mu^2})]+\frac{\mr{Tr}[U_L^tU_L^t+U_R^tU_R^t]}{2}-\mr{Tr}[U_L^t.M_u.U_R^t.M_u^{-1}]\nonumber\\
&+\mr{Tr}[\big(U_L^bU_L^b+U_R^bU_R^b+2Q_b(U_L^b+U_R^b)\big).(-\Delta_{\epsilon}+\log\frac{M_d^2}{\mu^2})]+\frac{\mr{Tr}[U_L^bU_L^b+U_R^bU_R^b]}{2}-\mr{Tr}[U_L^b.M_d.U_R^b.M_d^{-1}]\Big\}\nonumber\\
&=\frac{N_Cg^2}{32\pi^2c_W^2}\Big\{\frac{2}{3}-\frac{1}{3}\cos(2\theta_L^b)\cos(2\theta_R^b)-\frac{(m_b^2+m_B^2)\sin(2\theta_L^b)\sin(2\theta_R^b)}{6m_bm_B}-\frac{16}{9}[(s_L^t)^2\log\frac{m_X^2}{m_t^2}+(c_L^t)^2\log\frac{m_X^2}{m_T^2}]\nonumber\\
	&-\frac{5}{3}[(s_L^t)^2\log\frac{m_t^2}{m_bm_B}+(c_L^t)^2\log\frac{m_T^2}{m_bm_B}]-\frac{1}{9}\log\frac{m_t^2m_T^2}{m_b^2m_B^2}+\frac{7\cos(2\theta_L^b)+8\cos(2\theta_R^b)}{18}\log\frac{m_B^2}{m_b^2}\Big\}.
\end{align}
As we can see, the contributions from the non-$\chi_{\pm}$ part can not be simply described by the $\psi_{\pm}$ functions, which depend on the singlet and doublet properties. Then, the correct expression of $\Delta S^{XTB}$ can be calculated as follows:
\begin{align}\label{eqn:XTB:S}
&\Delta S^{XTB}=S_{non-\chi_{\pm}}^{XTB}+\frac{N_C}{2\pi}\Big\{-\psi_+(m_t^2,m_b^2)-(s_L^tc_L^t)^2\chi_+(m_t^2,m_T^2)\nonumber\\
&-[(s_L^bc_L^b)^2+4(s_R^bc_R^b)^2]\chi_+(m_b^2,m_B^2)-4(s_L^bc_L^b)(s_R^bc_R^b)\chi_-(m_b^2,m_B^2)\Big\}.
\end{align}

\subsection{Derivation of the $U$ parameter}
According to the Eq. \eqref{eqn:VLQ:dself}, the $U$ parameter also consists of the $\chi_{\pm}$ and non-$\chi_{\pm}$ parts. For the $\chi_{\pm}$ part, it can be calculated as
\begin{align}\label{eqn:XTB:Unonchi}
&\Delta U_{\chi_{\pm}}^{XTB}=-\frac{N_C}{2\pi}\Big\{2[(s_L^t)^2+(s_R^t)^2]\chi_+(m_X^2,m_t^2)+4s_L^ts_R^t\chi_-(m_X^2,m_t^2)\nonumber\\
&+2[(c_L^t)^2+(c_R^t)^2]\chi_+(m_X^2,m_T^2)+4c_L^tc_R^t\chi_-(m_X^2,m_T^2)\nonumber\\
&+[(c_L^tc_L^b+\sqrt{2}s_L^ts_L^b)^2+2(s_R^ts_R^b)^2-1]\chi_+(m_t^2,m_b^2)+2\sqrt{2}s_R^ts_R^b(c_L^tc_L^b+\sqrt{2}s_L^ts_L^b)\chi_-(m_t^2,m_b^2)\nonumber\\
&+[(c_L^ts_L^b-\sqrt{2}s_L^tc_L^b)^2+2(s_R^tc_R^b)^2]\chi_+(m_t^2,m_B^2)-2\sqrt{2}s_R^tc_R^b(c_L^ts_L^b-\sqrt{2}s_L^tc_L^b)\chi_-(m_t^2,m_B^2)\nonumber\\
&+[(s_L^tc_L^b-\sqrt{2}c_L^ts_L^b)^2+2(c_R^ts_R^b)^2]\chi_+(m_T^2,m_b^2)-2\sqrt{2}c_R^ts_R^b(s_L^tc_L^b-\sqrt{2}c_L^ts_L^b)\chi_-(m_T^2,m_b^2)\nonumber\\
&+[(s_L^ts_L^b+\sqrt{2}c_L^tc_L^b)^2+2(c_R^tc_R^b)^2]\chi_+(m_T^2,m_B^2)+2\sqrt{2}c_R^tc_R^b(s_L^ts_L^b+\sqrt{2}c_L^tc_L^b)\chi_-(m_T^2,m_B^2)\nonumber\\
&-(s_L^tc_L^t)^2\chi_+(m_t^2,m_T^2)-[(s_L^bc_L^b)^2+4(s_R^bc_R^b)^2]\chi_+(m_b^2,m_B^2)-4(s_L^bc_L^b)(s_R^bc_R^b)\chi_-(m_b^2,m_B^2)\Big\}.
\end{align}
For the non-$\chi_{\pm}$ part, it can be calculated as
\begin{align}
&\quad\frac{\alpha U_{non-\chi_{\pm}}^{XTB}}{4s_W^2}\nonumber\\
&=\frac{N_Cg^2}{96\pi^2}\Big\{[V_L^{Xt}(V_L^{Xt})^{\dag}-U_L^XU_L^X+V_R^{Xt}(V_R^{Xt})^{\dag}-U_R^XU_R^X](-\Delta_{\epsilon}+\log\frac{m_X^2}{\mu^2})\nonumber\\
&+\mr{Tr}[\big(V_L^{tb}(V_L^{tb})^{\dag}+(V_L^{Xt})^{\dag}V_L^{Xt}-U_L^tU_L^t+V_R^{tb}(V_R^{tb})^{\dag}+(V_R^{Xt})^{\dag}V_R^{Xt}-U_R^tU_R^t\big).(-\Delta_{\epsilon}+\log\frac{M_u^2}{\mu^2})]\nonumber\\
&+\mr{Tr}[\big((V_L^{tb})^{\dag}V_L^{tb}-U_L^bU_L^b+(V_R^{tb})^{\dag}V_R^{tb}-U_R^bU_R^b\big).(-\Delta_{\epsilon}+\log\frac{M_d^2}{\mu^2})]+V_L^{Xt}(V_L^{Xt})^{\dag}-\frac{1}{2}U_L^XU_L^X\nonumber\\
&+V_R^{Xt}(V_R^{Xt})^{\dag}-\frac{1}{2}U_R^XU_R^X+\mr{Tr}[V_L^{tb}(V_L^{tb})^{\dag}-\frac{1}{2}U_L^tU_L^t-\frac{1}{2}U_L^bU_L^b+V_R^{tb}(V_R^{tb})^{\dag}-\frac{1}{2}U_R^tU_R^t-\frac{1}{2}U_R^bU_R^b]\nonumber\\
&-\frac{1}{2}[V_L^{Xt}.M_u^{-1}.(V_R^{Xt})^{\dag}m_X+\frac{V_L^{Xt}.M_u.(V_R^{Xt})^{\dag}}{m_X}-U_L^XU_R^X+V_R^{Xt}.M_u^{-1}.(V_L^{Xt})^{\dag}m_X+\frac{V_R^{Xt}.M_u.(V_L^{Xt})^{\dag}}{m_X}-U_R^XU_L^X]\nonumber\\
&-\frac{1}{2}\mr{Tr}[V_L^{tb}.M_d^{-1}.(V_R^{tb})^{\dag}.M_u-U_L^b.M_d^{-1}.U_R^b.M_d+V_R^{tb}.M_d^{-1}.(V_L^{tb})^{\dag}.M_u-U_R^b.M_d^{-1}.U_L^b.M_d]\nonumber\\
&-\frac{1}{2}\mr{Tr}[V_L^{tb}.M_d.(V_R^{tb})^{\dag}.M_u^{-1}-U_L^t.M_u^{-1}.U_R^t.M_u+V_R^{tb}.M_d.(V_L^{tb})^{\dag}.M_u^{-1}-U_R^t.M_u^{-1}.U_L^t.M_u]\Big\}\nonumber\\
&=-\frac{N_Cg^2}{32\pi^2}\Big\{\frac{1}{3}-\frac{1}{3}\cos(2\theta_L^b)\cos(2\theta_R^b)-\frac{(m_b^2+m_B^2)\sin(2\theta_L^b)\sin(2\theta_R^b)}{6m_bm_B}-\frac{4}{3}[(s_L^t)^2+(s_R^t)^2]\log\frac{m_t^2}{m_X^2}\nonumber\\
	&-\frac{4}{3}[(c_L^t)^2+(c_R^t)^2]\log\frac{m_T^2}{m_X^2}+\frac{2}{3}[(s_L^b)^2+(s_R^b)^2]\log\frac{m_b^2}{m_X^2}+\frac{2}{3}[(c_L^b)^2+(c_R^b)^2]\log\frac{m_B^2}{m_X^2}\Big\}.
\end{align}
Note that the non-$\chi_{\pm}$ contributions vanish for the singlet and doublet VLQ \cite{Lavoura:1992np}, while it is non-zero for the $(X,T,B)_{L,R}$ triplet. Thus, the total contributions of $U$ parameter should be
\begin{align}\label{eqn:XTB:U}
&\Delta U^{XTB}=\Delta U_{\chi_{\pm}}^{XTB}+U_{non-\chi_{\pm}}^{XTB}.
\end{align}
\end{sloppypar}

\bibliography{mW-LQ-VLQ-CPC-v2, mW-anomalies-CPC-v2}
\end{document}